\documentclass{article}
\usepackage[margin=1in]{geometry}
\usepackage[utf8]{inputenc}
\usepackage{graphicx}
\usepackage{caption}
\usepackage{subcaption}
\usepackage{cite}
\usepackage{authblk}
\usepackage{multicol}
\usepackage{mathtools}
\usepackage{amsmath}
\usepackage{xcolor}
\usepackage{lineno}
\usepackage{amssymb}
\usepackage{hyperref}

\DeclareGraphicsExtensions{.pdf,.png,.jpg}
\graphicspath{ {figures/} }

\title{High $Q^2$ electron-proton elastic scattering at the future Electron-Ion Collider}

\author[1]{B.~Schmookler}
\author[1]{A.~Pierre-Louis}
\author[1,2]{A.~Deshpande}
\author[3]{D.~Higinbotham}
\author[4]{\\E.~Long}
\author[5]{A. J. R.~Puckett}

\affil[1]{Center for Frontiers in Nuclear Science and Stony Brook Univeristy}
\affil[2]{Brookhaven National Laboratory}
\affil[3]{Thomas Jefferson National Accelerator Facility}
\affil[4]{University of New Hampshire}
\affil[5]{University of Connecticut}

\date{\today}

\begin{document}

\maketitle

\begin{abstract}
Unpolarized electron-proton elastic scattering cross-section measurements at high $Q^2$ allow for improved extractions of the proton electromagnetic form factors as well as provide constraints on possible hard two-photon exchange effects. We present a detailed study of the feasibility of making these high $Q^2$ e-p elastic measurements at the future Electron-Ion Collider (EIC). The results show that e-p elastic cross sections can be obtained in the momentum transfer range of $6~(GeV/c)^2 < Q^2 < 40~(GeV/c)^2$, which would be the highest-ever $Q^2$ values measured. These data will all be at virtual photon polarizations close to unity, $\epsilon \sim 1$.
\end{abstract}

\section{Introduction} \label{sec:intro}
Electron-proton elastic scattering is a fundamental process in the study of nucleon structure. In the one-photon exchange approximation, the unpolarized e-p differential cross section can be written in the rest frame of the initial proton as

\begin{equation} \label{cs_rest}
    \frac{d\sigma}{d\Omega_{e}} = \sigma_{Mott} \frac{\tau G^2_M(Q^2) + \epsilon G^2_E(Q^2)}{\epsilon(1+\tau)} ~ . 
\end{equation}
The elastic cross section, $\frac{d\sigma}{d\Omega_{e}}$, is differential with respect to a single variable, chosen to be the scattered electron angle here. $\sigma_{Mott}$ represents the cross section for electrons scattering off a point-like target. $Q^2$ is the negative of the square of the four momentum of the exchanged photon, $\epsilon$ is the polarization of the virtual photon, and $\tau = \frac{Q^2}{4M_p^2}$. $G_E$ and $G_M$ are functions of $Q^2$ and represent the proton electric and magnetic form factors, respectively. These two form factors encapsulate the elastic structure of the proton.

The form factors in Eq.~\eqref{cs_rest} can be extracted at a given $Q^2$ by measurements at different values of $\epsilon$. This is referred to as the Rosenbluth separation technique. Fixed-target experiments demonstrated the expected linear $\epsilon$ dependence of the reduced cross section ($\tau G^2_M + \epsilon G^2_E$), as well as the scaling of the form factors, $ \mu_p G_E/G_M \approx 1 $. The reduced sensitivity of the reduced cross section to $G_E$ at higher $Q^2$ necessitated the development of polarization observables $-$ beam-target double-spin asymmetry measurements and recoil polarization measurements $-$ in order to make precise extractions of the form factors at higher $Q^2$. These polarization measurements, however, only extract the ratio of the form factors (i.e. $ \mu_p G_E/G_M$), not the individual form factors themselves. Surprisingly, the form-factor ratio extracted from these polarization measurements showed a strong decrease with increasing $Q^2$, in contrast to the scaling observed using the Rosenbluth separation method on unpolarized cross section measurements.

A possible explanation for this discrepancy is the presence of hard two-photon exchange effects in the elastic e-p cross section, which are not taken into account in the standard radiative corrections procedure~\cite{Guichon:2003qm}. Since this two-photon exchange effect and other possible $\epsilon$-dependent effects are not well understood, measurements of the unpolarized elastic cross section at new values in the $Q^2$-$\epsilon$ phase space are needed to provide important additional experimental insight into the origin of the form-factor ratio discrepancy.

The future Electron-Ion Collider (EIC) will collide polarized electrons with polarized protons and light nuclei, as well as unpolarized heavy nuclei, at variable center-of-mass energies ($\sqrt{s} = 28 - 141$~GeV per nucleon) and high luminosity ($10^{33} - 10^{34}$~cm$^{-2}$~s$^{^-1}$). The EIC will therefore make important contributions to our knowledge of the underlying partonic structure of matter through Deep Inelastic Scattering (DIS), including the systematic study of the potential onset of non-linear QCD dynamics at high energy (low x)  and a better understanding of the origin of the nucleon spin.

In addition to these DIS measurements, the EIC, as we will demonstrate in this paper, has the ability to measure the elastic e-p cross section in the range $6~(GeV/c)^2 < Q^2 < 40~(GeV/c)^2$. This would be the highest $Q^2$ ever measured for e-p elastic scattering, and it would be the first time the elastic cross section is measured at a high-energy collider. Using these new measurements in global analyses would provide additional constraints on potential $\epsilon$-dependent effects~\cite{PhysRevC.76.035205}. Moreover, the over-constrained kinematics of elastic scattering would allow these events to be used for important detector calibration purposes.

A study of possible asymmetry measurements using polarized elastic e-p scattering at a high-energy collider for lower values of momentum transfer ($Q^2<3~(GeV/c)^2$) was performed in Ref.~\cite{Sofiatti:2011yi}. Here, we consider unpolarized elastic e-p scattering at higher $Q^2$. This paper is organized as follows: 

Section~\ref{sec:kinematics} discusses the expected statistical precision of the cross section measurements at EIC energies, as well as the kinematics of the outgoing electron and proton. A simple Born-level elastic event generator, which is used for reconstruction and background studies, is introduced here.

Section~\ref{sec:detector} describes an acceptance and resolution parameterization for a possible EIC central detector based on the recent EIC Yellow Report~\cite{AbdulKhalek:2021gbh}. This detector parameterization is implemented using the EIC-Smear~\cite{eicsmear} software package.

Section~\ref{sec:reconstruction} shows the ability of the central detector to reconstruct both the $Q^2$ and the proton mass peak for elastic e-p events. Elastic events are first generated and then reconstructed using the fast simulation model of the central detector discussed above. We emphasize that to adequately reconstruct the elastic events, both the electron and proton have to be successfully measured in the central detector.

Section~\ref{sec:bg} addresses the question of how well the high-$Q^2$ elastic events can be separated from the inelastic background. We use the \textit{Pythia6}~\cite{Sjostrand:2006za} event generator to create both an inclusive high-$Q^2$ inelastic sample as well as a sample based on the raw electron-proton spectrum (a minimum-bias sample), and then pass these events through the same fast detector simulation as the elastic events. We construct a set of elastic selection criteria and determine how much of the inelastic background survives. We also briefly consider the question of particle identification for the elastic electron and proton.

Finally, section~\ref{sec:concl} presents our conclusions and a discussion on potential future studies. (The EIC will have a non-zero beam crossing angle, and we consider the overall effects of this on the elastic production rates and kinematics in Appendix \ref{sec:app_ca}.)

\section{Elastic production rates and kinematics} \label{sec:kinematics}
We first calculate the expected elastic scattering yields and the kinematics of the outgoing electron and proton. We do these calculations for the EIC beam energy settings studied in the 2021 EIC Conceptual Design Report (CDR)~\cite{EIC_cdr} $-$ 5~GeV electrons on 41~GeV protons; 5~GeV electrons on 100~GeV protons; 10~GeV electrons on 100~GeV protons; and 18~GeV electrons on 275~GeV protons. In this work, we assume the incoming proton is along the positive z direction and the incoming electron is along the negative z direction. (A study on the impact of a non-zero crossing angle is presented in Appendix~\ref{sec:app_ca}.) We present our results for an integrated luminosity of 100~fb$^{-1}$, which corresponds to 230 days of running at $10^{34}$~cm$^{-2}$~s$^{^-1}$ assuming 50\% running efficiency.

For calculating expected event yields in a collider setting, we first recast Eq.~\eqref{cs_rest} in a Lorentz-invariant form as
\begin{equation} \label{cs_lorentz}
\frac{d\sigma}{dQ^2} = \frac{4\pi\alpha^2}{Q^4}\left[ \frac{G_E^2+\tau G_M^2}{1+\tau}\left(1-y-\frac{M_p^2y^2}{Q^2}\right)+\frac{1}{2}y^2G_M^2 \right]~,
\end{equation}
where the inelaticity variable y is equal to the fraction of the initial electron's energy that is lost during the interaction in the rest frame of the proton, $\alpha$ is the fine-structure constant, and $M_p$ is the mass of the proton.

\begin{figure}[ht]
\centering
\includegraphics[width=0.75\textwidth]{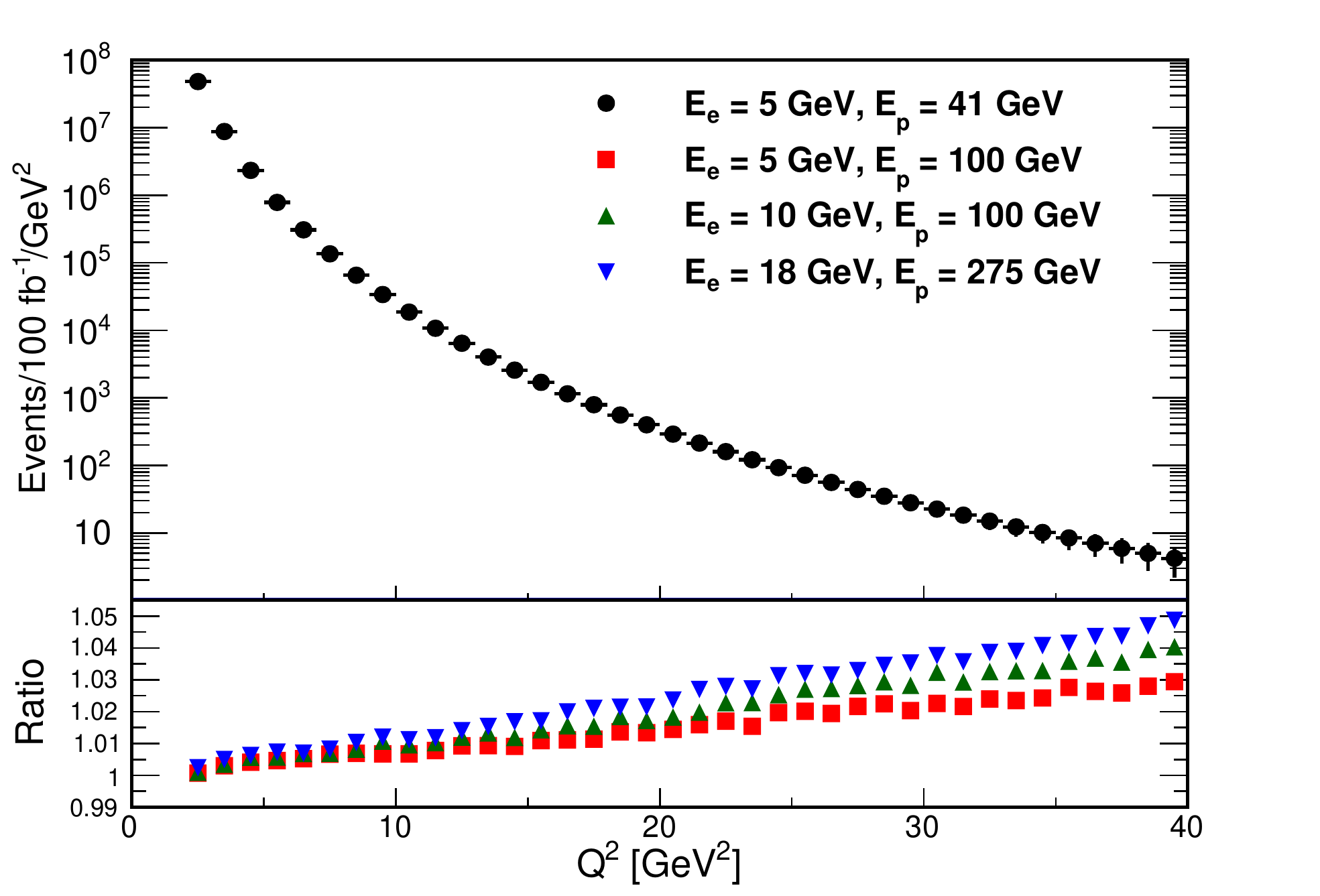}
\caption{Top panel: expected number of elastic events vs. $Q^2$ for 100 fb$^{-1}$ integrated luminosity and $Q^2$ bins of width 1~$(GeV/c)^2$. The result, which is calculated using Eq.~\eqref{cs_lorentz}, is shown for a 5 GeV electron beam on a 41 GeV proton beam. The expected event yield falls sharply with increasing $Q^2$. Bottom panel: ratio of the cross section for the other three EIC energies to the cross section for the 5 GeV electron on 41 GeV proton setting. Within statistical uncertainties, the cross section is independent of the beam energies at the EIC.}
\label{fig:events_one}
\end{figure}

Figure~\ref{fig:events_one} shows the expected number of elastic events for 100~fb$^{-1}$ integrated luminosity in $Q^2$ bins of width $1~(GeV/c)^2$ for a 5 GeV electron beam on a 41 GeV proton beam, with vertical error bars indicating the statistical uncertainty. The bottom panel in the figure shows the ratio of the cross section for the other three EIC energies to the cross section for the 5 GeV electron on 41 GeV proton setting. The form factors are parameterized using the functional form of Ref.~\cite{Kelly:2004hm}, with the fit parameters taken from Ref.~\cite{Puckett:2010kn}. The event counts for this luminosity and binning are approximately 1~million at $Q^2 = 5~(GeV/c)^2$, approximately 200 at $Q^2 = 20~(GeV/c)^2$, and approximately 5 at $Q^2 = 40~(GeV/c)^2$. Several years of high-luminosity data-taking, as well as combining bins at higher $Q^2$, would allow for better statistical precision.

An interesting feature shown in the bottom panel of figure~\ref{fig:events_one} is the lack of dependence of the cross section on the beam energy setting. This occurs because the inelasticity, y, is very small for elastic events at the EIC with $Q^2<40~(GeV/c)^2$. For a high-energy collider like the EIC, where the masses of the incoming electron and proton can be ignored (the Bj{\"o}rken limit), $Q^2$, $y$, the Bj{\"o}rken scaling variable $x$, and the square of the center of mass energy $s$ are related by the equation $Q^2 = sxy$. $s = 4E_eE_p$, where $E_e$ and $E_p$ are the energies of the incoming electron and proton, respectively. For elastic events $x=1$, which means that y reaches a maximum of 0.05 for $Q^2 < 40~(GeV/c)^2$. So at the EIC, Eq.~\eqref{cs_lorentz} reduces to approximately
\begin{equation}
\frac{d\sigma}{dQ^2} \approx \frac{4\pi\alpha^2}{Q^4}\left[ \frac{G_E^2+\tau G_M^2}{1+\tau} \right]~.
\end{equation}

The polarization of the virtual photon, $\epsilon$, can be written in a Lorentz-invariant form as~\cite{HERMES:2004zsh}
\begin{equation}
\epsilon = \frac{1-y-\frac{1}{4}\gamma^2y^2}{1-y+\frac{1}{4}y^2\left(\gamma^2+2\right)}~,
\end{equation}
where $\gamma^2 = \frac{Q^2}{\nu^2}$. $\nu$ is the energy of the virtual photon in the incoming proton rest frame, and for elastic scattering is equal to $\frac{Q^2}{2M_p}$. As y is always less than 0.05 and $y^2\gamma^2 \ll 1$, the virtual photon is fully polarized ($\epsilon \sim 1$) for all values of $Q^2$ accessible at the EIC.

The angle of the outgoing electron ($\theta_e$), the energy of the outgoing electron ($E'_e$), the angle of the outgoing proton ($\theta_p$), and the energy of the outgoing proton ($E'_p$) can be written in terms of the kinematic variable y (or $Q^2$) and the energies of the incoming electron ($E_e$) and incoming proton ($E_p$). In the Bj{\"o}rken limit, these equations are~\cite{Wolf:1994es}
\begin{align}
\begin{split}
E'_e & = (1-y)E_e + yE_p~, \\
\cos\theta_e & = \frac{yE_p - (1-y)E_e}{yE_p+(1-y)E_e}~, \\
E'_p & = yE_e + (1-y)E_p~, \\
\cos\theta_p & = \frac{-yE_e+(1-y)E_p}{yE_e+(1-y)E_p}~.
\end{split}
\end{align}

\begin{figure}[ht]	
	\centering
	\begin{subfigure}[b]{0.49\textwidth}
	    \centering
         \includegraphics[keepaspectratio=true,width=2.9in,page=2]{kin_yellow_proton}	
	\end{subfigure}
	\hfill
	\begin{subfigure}[b]{0.49\textwidth}
	    \centering
		\includegraphics[keepaspectratio=true,width=2.9in,page=3]{kin_yellow_proton}
	\end{subfigure}
    \begin{subfigure}[b]{0.49\textwidth}
		\centering
		\includegraphics[keepaspectratio=true,width=2.9in,page=1]{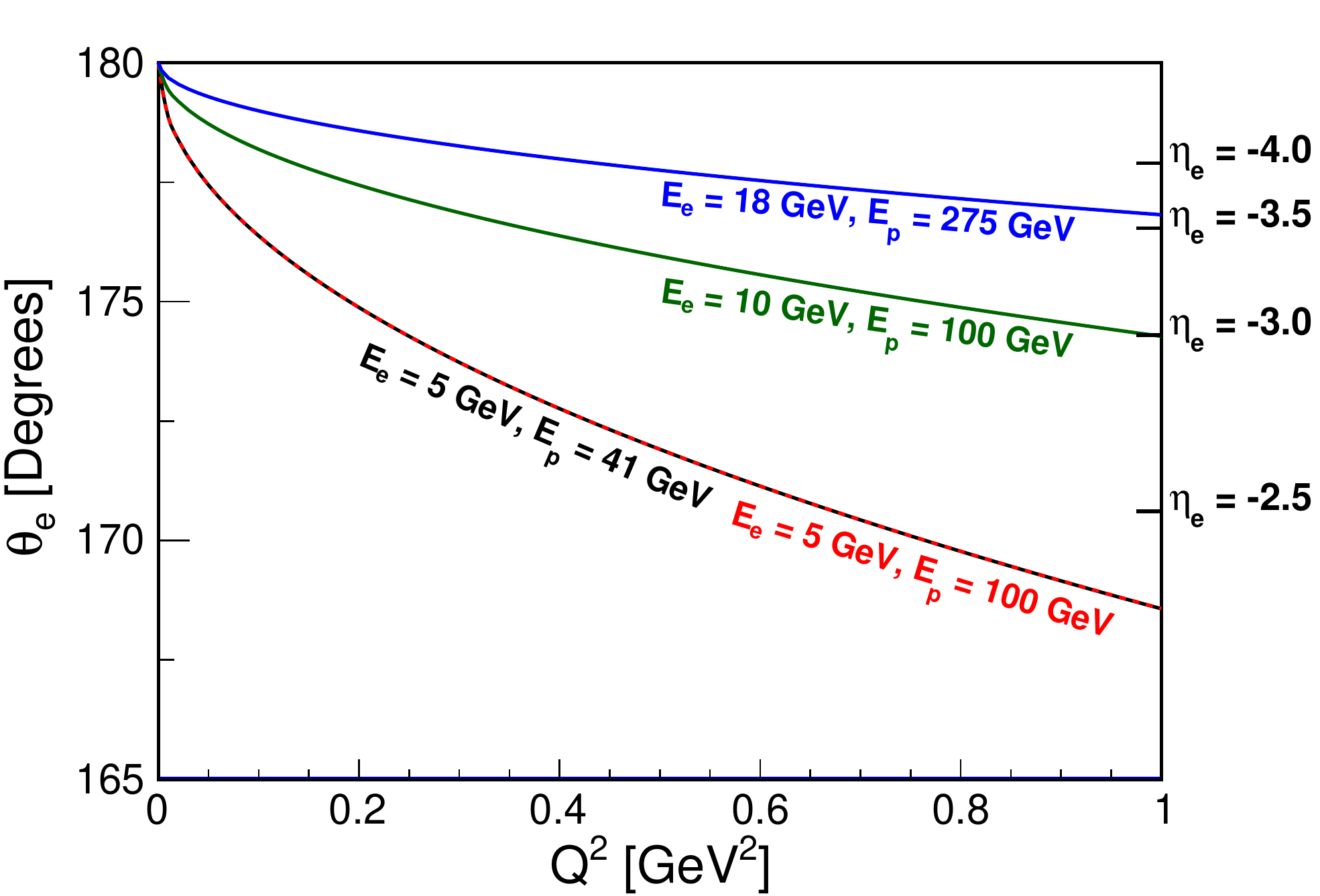}
	\end{subfigure}
	\hfill
	\begin{subfigure}[b]{0.49\textwidth}
		\centering
		\includegraphics[keepaspectratio=true,width=2.9in,page=4]{kin_yellow_proton}
	\end{subfigure}
	
	\caption{Top left: elastically scattered electron angle vs. $Q^2$; bottom left: electron angle vs. $Q^2$ for $Q^2 < 1~(GeV/c)^2$; top right: elastically scattered proton angle vs. $Q^2$; bottom right: proton angle vs. $Q^2$ for $Q^2 < 10~(GeV/c)^2$. All angles are measured with respect to the incoming proton beam direction. Several corresponding pseudo-rapidity $\eta$) values are given on the right vertical axis of each plot. For a central EIC detector with acceptance $-3.5 < \eta < 3.5$, the proton will only be within the acceptance for the lowest beam energy setting.} \label{fig:events_two}
\end{figure}

Using the above set of equations, the angles for the outgoing electron and proton are plotted as a function of $Q^2$ in figure~\ref{fig:events_two}. The scattering angle for both the electron and proton are measured with respect to the incoming proton beam. As can be seen, the scattering angle of the electron decreases with increasing $Q^2$, while the angle of the proton increases with increasing $Q^2$. For an EIC central detector acceptance that covers the pseudo-rapidity range of [-3.5, 3.5] (see section~\ref{sec:detector} for details), the scattered electron will be within the acceptance for $Q^2 \gtrsim 1~(GeV/c)^2$ for all energy settings. The scattered proton, however, will only reach the central detector for the lowest beam energy combination $-$ 5 GeV electrons on 41 GeV protons $-$ and, at this setting, for $Q^2 \gtrsim 6~(GeV/c)^2$ only. As we will see in section~\ref{sec:reconstruction}, both the electron and proton need to be reconstructed to correctly measure the elastic events. For the higher $\sqrt{s}$ settings, the elastic proton may be detected in the far-forward part of the detector, but we do not study this possibility in this work.

\begin{figure}[ht]	
	\centering
	\begin{subfigure}[b]{0.49\textwidth}
	    \centering
         \includegraphics[keepaspectratio=true,width=2.9in,page=5]{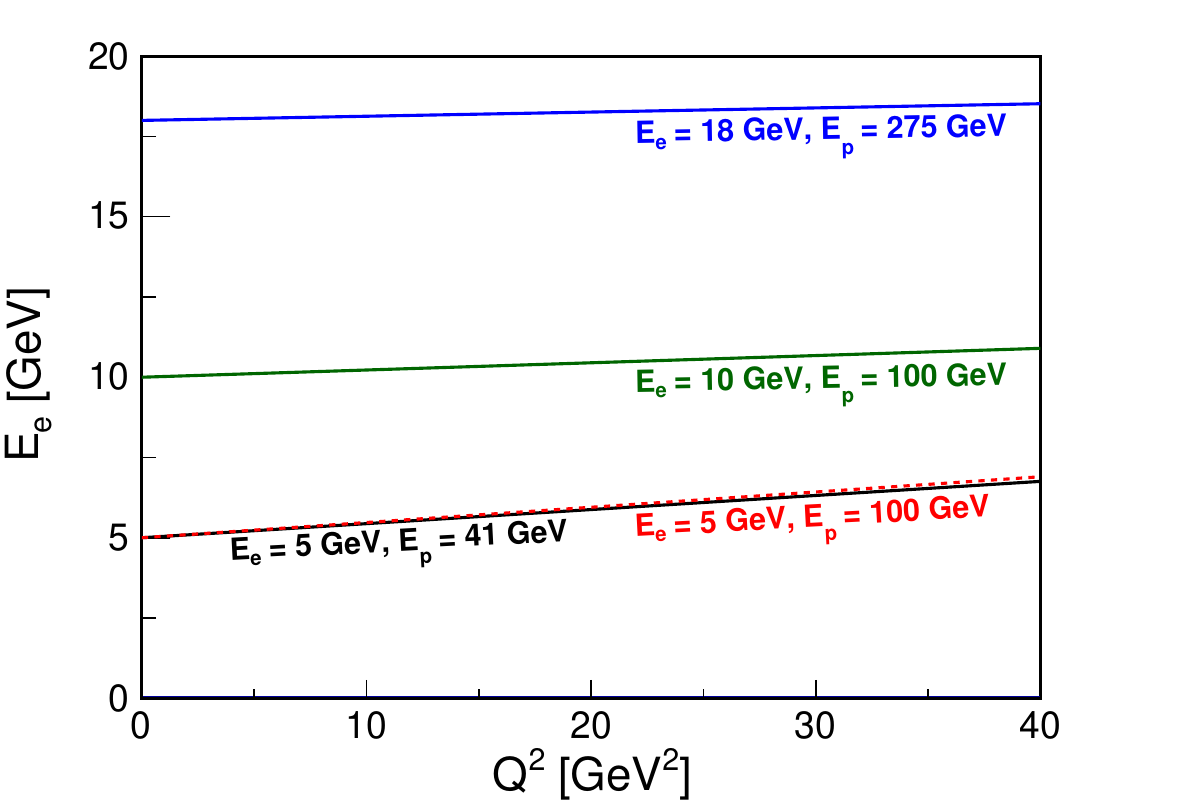}
	\end{subfigure}
	\hfill
	\begin{subfigure}[b]{0.49\textwidth}
	    \centering
		\includegraphics[keepaspectratio=true,width=2.9in,page=6]{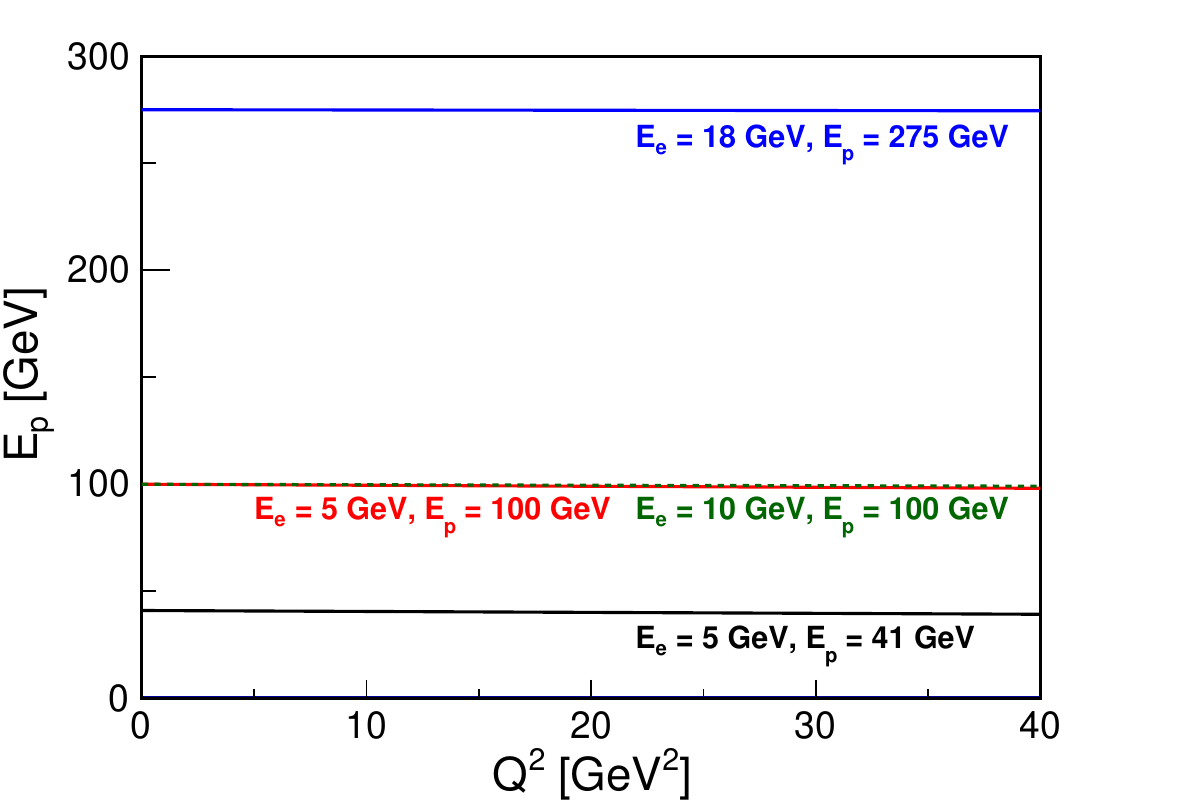}
	\end{subfigure}
	\caption{Left: elastically scattered electron energy vs. $Q^2$; right: elastically scattered proton energy vs. $Q^2$. The energy of the electron (proton) increases (decreases) gradually from the beam energy with increasing $Q^2$. Therefore, the elastic electron will always have a large enough transverse momentum to reach detectors beyond the central tracking detector. Similarly, the outgoing proton for the 41~GeV energy setting will also have enough momentum to reach the particle identification detectors within the central detector.} \label{fig:events_three}
\end{figure}

The energies of the outgoing electron and proton are plotted in figure~\ref{fig:events_three}. The scattered electron energy increases gradually from the beam energy with increasing $Q^2$, and the scattered proton energy decreases gradually with increasing $Q^2$. Figures~\ref{fig:events_two} and \ref{fig:events_three} show that at the EIC the kinematics of the outgoing elastic electron (proton) are approximately independent of the proton (electron) beam energy.

\begin{figure}[ht]
\centering
\includegraphics[width=0.75\textwidth]{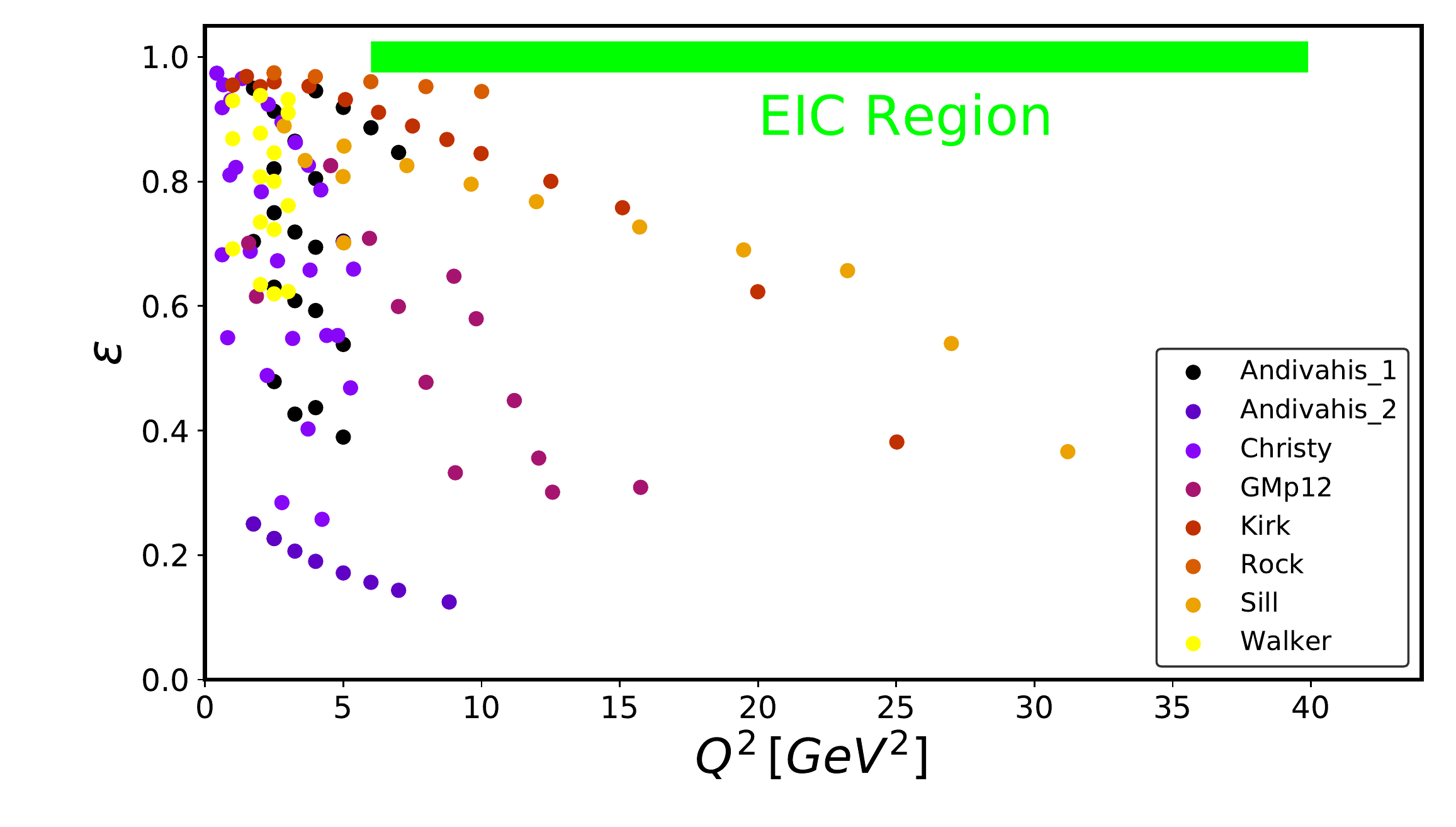}
\caption{Expected elastic e-p EIC coverage in the $Q^2-\epsilon$ phase space compared with existing high-$Q^2$ elastic cross section measurements. Data are from Refs.~\cite{Andivahis:1994rq,E94110:2004lsx,Christy:2021snt,Kirk:1972xm,Rock:1991jy,Sill:1992qw,PhysRevD.49.5671}. As can be easily seen, the EIC will access a part of the kinematic phase space that fixed-target experiments have been unable to study.}
\label{fig:events_four}
\end{figure}

With the above information on the expected yields and kinematics, we can conclude that the EIC will be able to make measurements of the elastic cross section in the range $6~(GeV/c)^2 < Q^2 < 40~(GeV/c)^2$ at $\epsilon \sim 1$. This EIC coverage is shown in figure~\ref{fig:events_four} along with the existing high-$Q^2$ fixed-target data. It is apparent that the EIC will access a wide area in the $Q^2$-$\epsilon$ phase space where no data currently exist.

We next created a simple Born-level elastic generator using the \textit{Foam} library~\cite{Jadach:2002kn} incorporated into the ROOT software framework~\cite{BRUN199781}. For each of the four EIC energy settings, the energy of the incoming electron in the rest frame of the proton is calculated. Events are then generated in the rest frame of the incoming proton according to Eq.~\eqref{cs_rest}, with the form factors again parameterized based on Ref.~\cite{Kelly:2004hm} and Ref.~\cite{Puckett:2010kn}. Once the angle of the scattered electron is known in the incoming proton's rest frame, the energy of the scattered electron and both the angle and energy of the scattered proton can be directly calculated. The event is then Lorentz transformed back into the collider frame.

For each of the four beam energy combinations, we generated 100 fb$^{-1}$ worth of elastic data with $Q^2 > 5~(GeV/c)^2$. This corresponds to approximately 1.4 million events per energy setting. We confirmed that the generator reproduced the calculations shown in figures~\ref{fig:events_one}-\ref{fig:events_three}. In the following sections, we use these generated events to study the identification and reconstruction of elastic scattering at the EIC.

\clearpage

\section{Detector Parameterization} \label{sec:detector}
In order study the reconstruction of the high-$Q^2$ elastic events (section~\ref{sec:reconstruction}), as well as the ability separate these events from the inelastic background (section~\ref{sec:bg}), we make use of a parameterization of an EIC central detector based primarily on studies done in the recent EIC Yellow Report~\cite{AbdulKhalek:2021gbh}.

\begin{figure}[ht]
\centering
\includegraphics[width=0.75\textwidth]{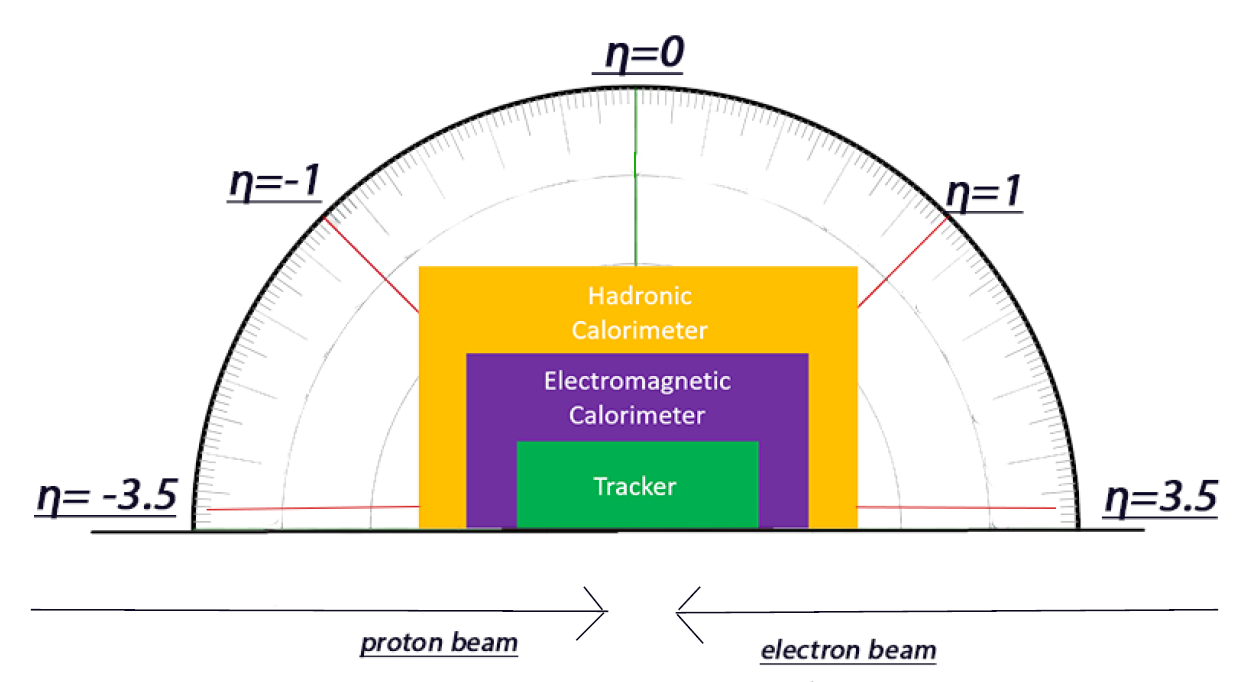}
\caption{Cartoon showing the relative locations of the EIC central detector subsystems relevant to this work $-$ the tracking detector, the electromagnetic calorimeter, and the hadronic calorimeter~\cite{protractor}. The directions of the electron and proton beams are shown, as well as lines indicating several values of pseudo-rapidity. The sizes of the included detectors are not drawn to scale.}
\label{fig:detector_one}
\end{figure}

We define the EIC central detector as comprising all detector subsystems contained within the electron end-cap (negative rapidity), the barrel region (mid-rapidity), and the hadron end-cap (forward rapidity). The relevant detectors for this work are the tracking detector, the electromagnetic calorimeter, and the hadronic calorimeter. A cartoon of these three detector systems is shown in figure~\ref{fig:detector_one}. All detectors are assumed to have an pseudo-rapidity (angular) acceptance extending from $\eta = -3.5$ to $\eta = +3.5$, with hermetic coverage in the azimuthal direction. No minimum transverse momentum acceptance limit is assumed is this work, but, as discussed above, the elastic electron and proton have high enough momentum to reach the outer detectors for any reasonable central detector magnetic field. 

The tracking detector reconstructs the momentum, polar angle, and azimuthal angle for all charged particles. The resolutions used for the tracking detector are shown in table~\ref{tab:tracker}. The electron energy is reconstructed in the electromagnetic calorimeter and proton energy is reconstructed in the hadronic calorimeter. The resolutions for these two calorimeters are shown in table~\ref{tab:calo}.

The resolutions shown in tables~\ref{tab:tracker} and \ref{tab:calo} are implemented in the Eic-Smear software framework. For every generated event and particle, the true quantities $-$ momentum, polar angle, azimuthal angle, and energy $-$ are smeared based on a Gaussian function with the standard deviations given in table~\ref{tab:tracker} and \ref{tab:calo}.

In the work that follows, we assume that the electron and proton have been correctly identified by their respective particle identification (PID) detectors. (We will briefly discuss PID questions in section~\ref{sec:bg}). Once a charge particle has been correctly identified, the question arises of how to perform the momentum and energy reconstruction in the most optimal way. For the electron, the tracking detector will have better resolution at lower energy, while the electromagnetic calorimeter will be superior at higher energies. For simplicity, we choose to always reconstruct the electron energy using the calorimeter, and then calculate the momentum based on this energy and the knowledge that the particle is an electron. For the elastic proton, where only the 41~GeV proton beam is relevant, the tracker is always superior to the hadronic calorimeter. So, we reconstruct the proton momentum using the tracker and then calculate the proton's energy from this momentum and the knowledge that the particle is a proton. For both the electron and the proton, the angles are reconstructed using the resolutions given in table~\ref{tab:tracker}.

\begin{table}[ht]
\centering
\begin{tabular}
{c||c||c||c}
$\eta$ range & $\sigma_{p}/p \thinspace [\%]$ & $\sigma_{\theta}$ \thinspace [Rad] &
$\sigma_{\phi} \thinspace [Rad] $\\
\hline
-3.5 $-$ -2.0  & $ 0.1 \cdot p \bigoplus 0.5 $ & &\\

-2.0 $-$ -1.0  & $ 0.05 \cdot p \bigoplus 0.5 $ & & \\

-1.0 $-$ +1.0  & $ 0.05 \cdot p \bigoplus 0.5 $ & $0.01 / \left(p \cdot \sqrt{\sin\theta}\right)$ & $0.01$ \\

+1.0 $-$ +2.5  & $ 0.05 \cdot p \bigoplus 1.0 $ & & \\

+2.5 $-$ +3.5  & $ 0.1 \cdot p \bigoplus 2.0 $ & &
\end{tabular} 
\caption{Tracking detector parameterization of the momentum, polar angle, and azimuthal angle for charged particles. The first column shows the pseudo-rapidity range; the second the momentum resolution; the third the polar angle resolution; and the fourth the azimuthal angle resolution. The $\bigoplus$ symbol indicates that the quantities are combined in quadrature. The resolution for these quantities is assumed to be independent of the particle species.}
\label{tab:tracker}
\end{table}

\begin{table}[ht]
\centering
\begin{tabular}
{c||c||c}
$\eta$ range & EM Calorimeters $\sigma_{E}/E \thinspace [\%]$ &  Hadronic Calorimeters $\sigma_{E}/E \thinspace [\%]$ \\
\hline
 -3.5 $-$ -2.0  & $2/\sqrt{E}$ & $50/\sqrt{E}$ \\
 -2.0 $-$ -1.0  & $7/\sqrt{E}$ & $50/\sqrt{E}$ \\
 -1.0 $-$ +1.0  & $12/\sqrt{E}$& $85/\sqrt{E} \bigoplus 7$\\
 +1.0 $-$ +3.5  & $12/\sqrt{E}$& $50/\sqrt{E}$
\end{tabular} 
\caption{Calorimeter detector parameterizations of the energy resolution. The first column shows the pseudo-rapidity range; the second column shows the electromagnetic calorimeter energy resolution which is used for electrons, positrons, and photons; and the hadronic calorimeter energy resolution which is used for protons and other hadronic particles. The $\bigoplus$ symbol indicates that the quantities are combined in quadrature.}
\label{tab:calo}
\end{table}

\section{Reconstruction} \label{sec:reconstruction}
For a given incoming electron and proton beam energy combination, two variables need to be reconstructed to characterize whether the event is an elastic scattering event. We need to measure $Q^2$ as well as a variable to determine whether the event is near the elastic peak. The latter is done by comparing the reconstructed hadronic final-state mass (W) to the proton mass, which means requiring either $W \sim M_p$ or, equivalently, the Bj{\"o}rken scaling variable $x \sim 1$. We will focus on the reconstruction of $x$ here. The elastic cross section at a given $Q^2$ is determined by integrating events over the elastic peak, allowing for soft real photon radiation only.

The required kinematic variables can be reconstructed using the scattered electron. This is, of course, the method that is primarily used in fixed-target experiments. We can relate the variables $y$, $Q^2$ and $x$ to the energy and polar angle (defined as above with respect to the incoming proton direction) of the scattered electron as follows~\cite{KLEIN2008343}:

\begin{align}
\begin{split}
y_{e} &= 1 - \frac{E'_e}{2E_e}(1 - \cos\theta_e)~, \\
Q^2_e &= 4E_eE'_e\cos^2(\theta_{e}/2)~, \\
x_e &= \frac{Q^2_e}{s y_e}~.
\end{split}
\end{align}

\begin{figure}[ht]	
	\centering
	\begin{subfigure}[b]{0.49\textwidth}
	    \centering
         \includegraphics[keepaspectratio=true,width=2.9in,page=7]{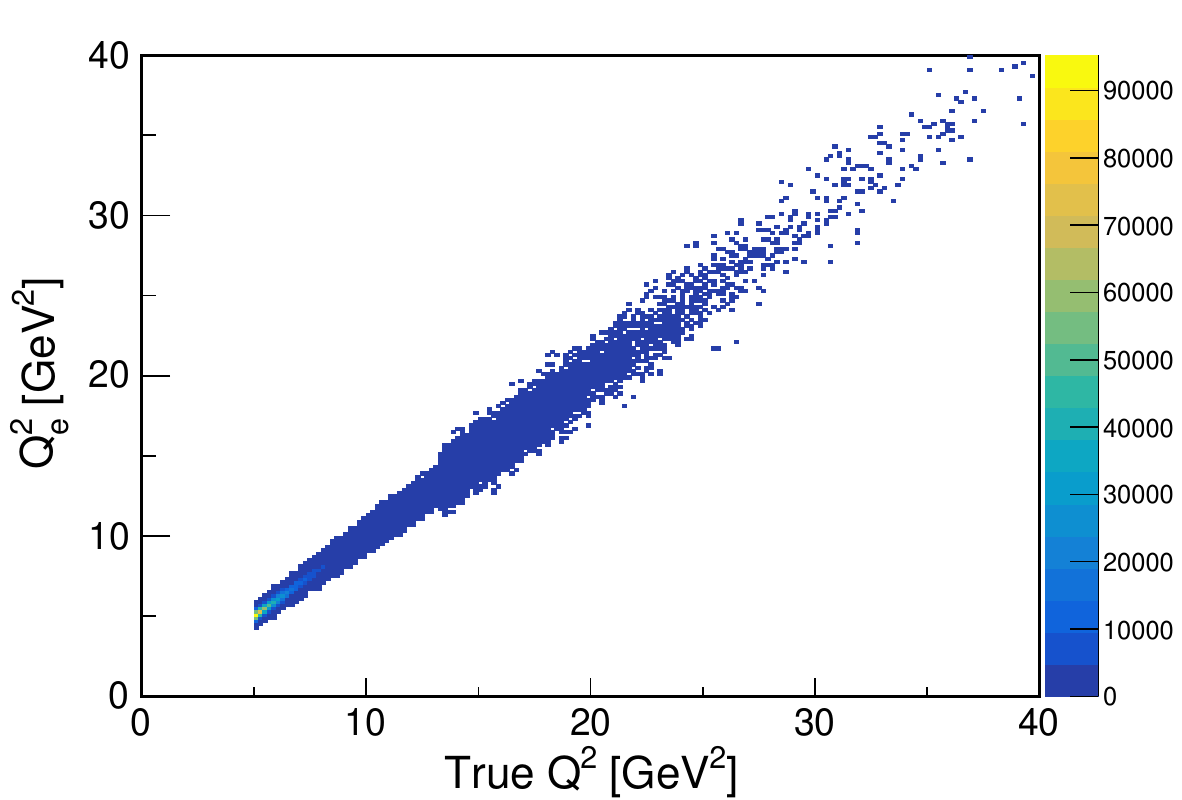}
		\caption{5~GeV e$^-$ x 41~GeV p} \label{fig:reconstruction_one_a}
	\end{subfigure}
	\hfill
	\begin{subfigure}[b]{0.49\textwidth}
	    \centering
		\includegraphics[keepaspectratio=true,width=2.9in,page=8]{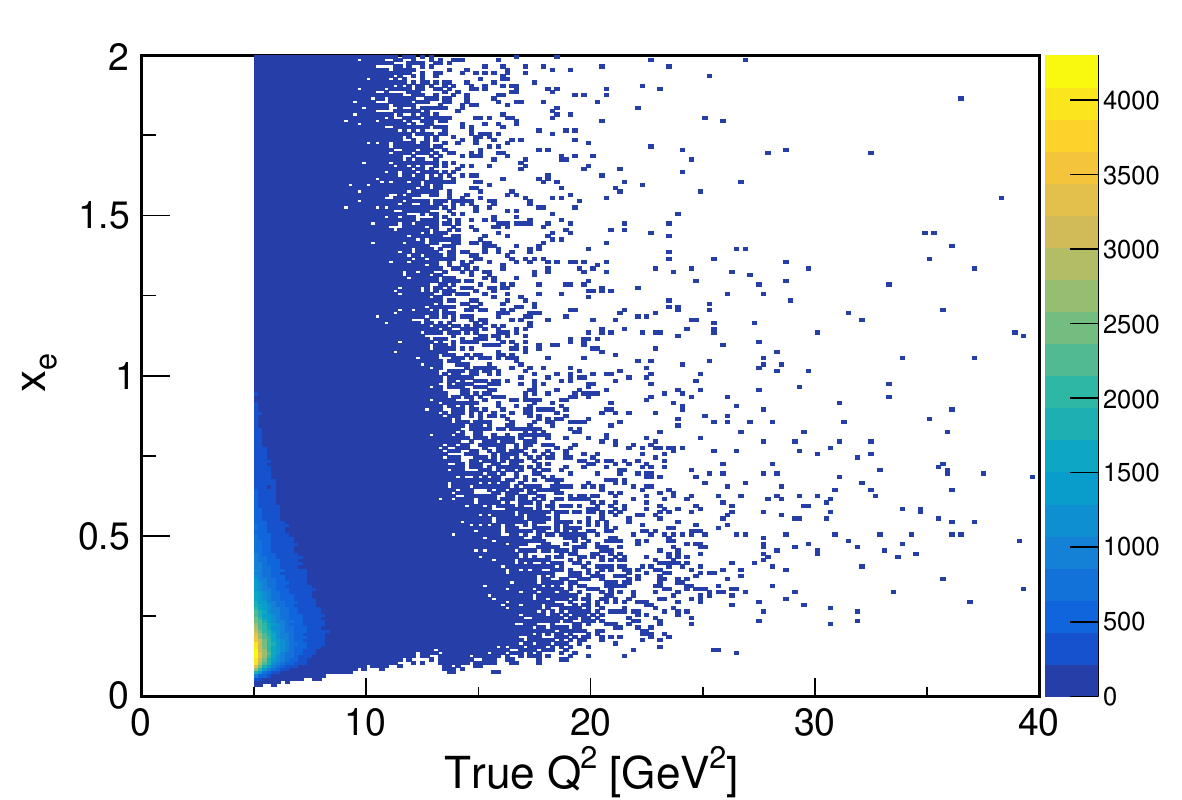}
		\caption{5~GeV e$^-$ x 41~GeV p} \label{fig:reconstruction_one_b}
	\end{subfigure}
	\hfill
	\begin{subfigure}[b]{0.49\textwidth}
	    \centering
         \includegraphics[keepaspectratio=true,width=2.9in,page=7]{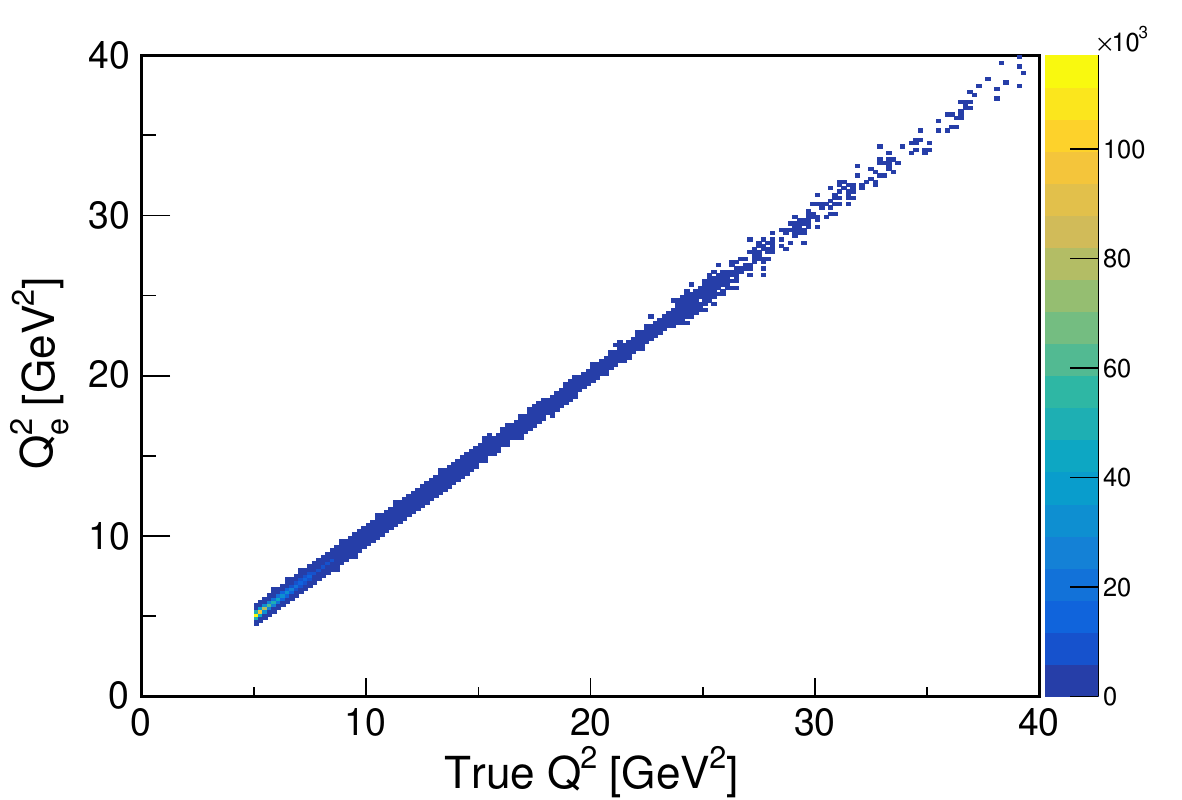}
		\caption{18~GeV e$^-$ x 275~GeV p} \label{fig:reconstruction_one_c}
	\end{subfigure}
	\hfill
	\begin{subfigure}[b]{0.49\textwidth}
	    \centering
		\includegraphics[keepaspectratio=true,width=2.9in,page=8]{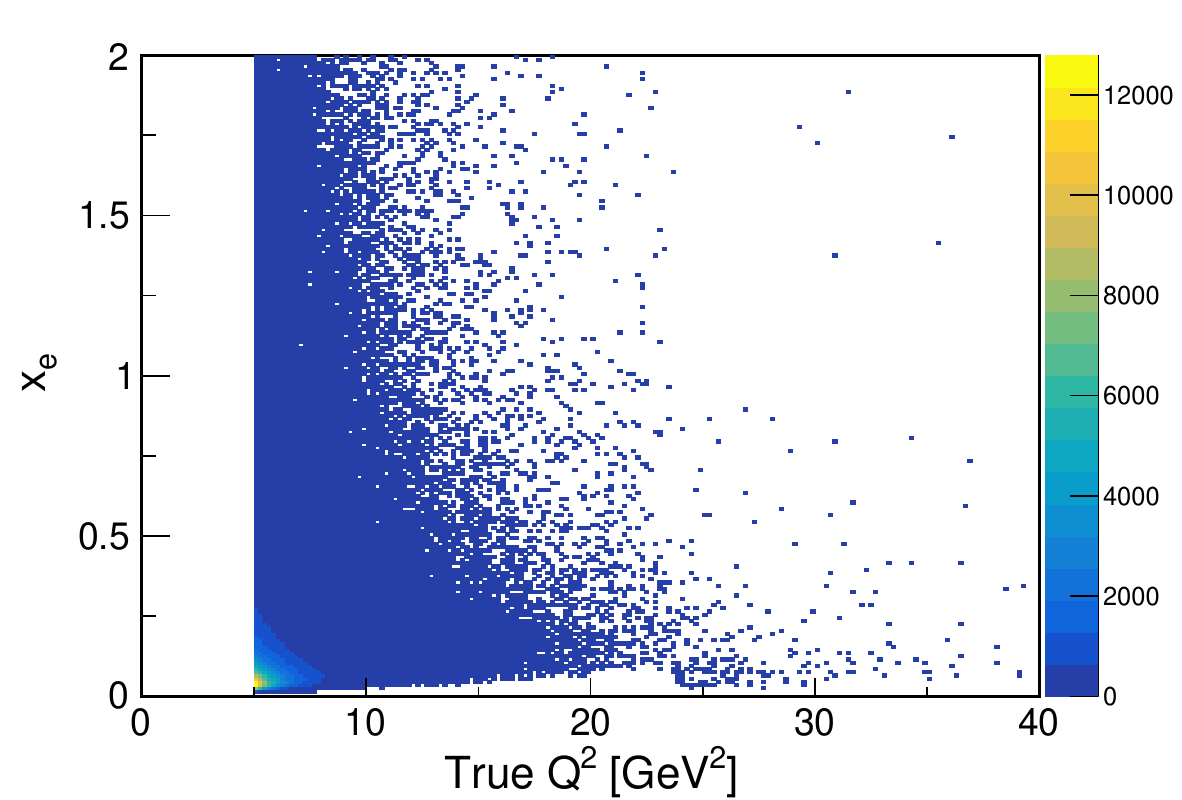}
		\caption{18~GeV e$^-$ x 275~GeV p} \label{fig:reconstruction_one_d}
	\end{subfigure}
	\caption{Reconstruction of the kinematic variables relevant to elastic e-p scattering using the scattered electron method as a function of the true $Q^2$. Top left: reconstruction of $Q^2_e$ for the lowest $\sqrt{s}$ setting; top right: reconstruction of $x_e$ for the lowest $\sqrt{s}$ setting; bottom left: reconstruction of $Q^2_e$ for the highest $\sqrt{s}$ setting; bottom right: reconstruction of $x_e$ for the highest $\sqrt{s}$ setting. The events are generated using the Born-level elastic generator introduced in section~\ref{sec:kinematics}. The elastic peak is should be located at $x_e = 1$, but it is very poorly reconstructed using the scattered electron method.} \label{fig:reconstruction_one}
\end{figure}

Figure~\ref{fig:reconstruction_one} shows the reconstructed $Q_e^2$ and $x_e$ for the generated elastic events vs. the true $Q^2$ using the electron reconstruction method. This is shown for both the 5~GeV electron on 41~GeV and the 18~GeV electron on 275~GeV proton energy combinations. The reconstructed $Q_e^2$ is reasonable and improves for the higher energy electron beam $-$ where the elastically scattered electron has a higher energy and the electromagenetic calorimeter energy resolution is better. For elastic events, the Bj{\"o}rken scaling variable should be reconstructed at exactly $x=1$. However, as can be clearly seen in figures~\ref{fig:reconstruction_one_b} and \ref{fig:reconstruction_one_d}, the reconstruction of the elastic peak using the standard electron method is impossible for all EIC beam energy settings and the entire $Q^2$ range simulated.

The reason for this poor reconstruction of $x_e$ is the inverse dependence of the $x_e$ resolution on $y_e$~\cite{KLEIN2008343}:
\begin{equation}
\frac{\delta x_e}{x_e} = \frac{1}{y_e} \cdot \frac{\delta E'_e}{E'_e} \bigoplus \left[ \frac{1-y_e}{y_e} \cdot \cot{\frac{\theta_e}{2}} +\tan{\frac{\theta_e}{2}} \right] \cdot \delta \theta_e~.
\end{equation}
Since elastic scattering events at the EIC for $Q^2 < 40~(GeV/c)^2$ will be at very low y, the $x_e$ resolution will be extremely poor for any realistic detector resolution.

The solution for accurate reconstruction of $Q^2$ and $x$ for elastic events is to use information from the elastically scattered proton. We consider two reconstruction methods: 1) the application the Jacquet-Blondel (JB) method to elastic scattering, which means reconstructing the kinematic variables using only the outgoing proton, and 2) the Double-Angle (DA) method, which uses the polar angles of both the electron and proton for reconstruction. For the JB method, $Q^2$ and $x$ can be related to energy and momentum vector of the outgoing proton as~\cite{blondel,KLEIN2008343}
\begin{align}
\begin{split}
y_{JB} &= \frac{E'_p - p_{z,p}}{2E_e}~, \\
Q^2_{JB} &= \frac{p_{T,p}^2}{1 - y_{JB}}~ \\
x_{JB} &= \frac{Q^2_{JB}}{s y_{JB}}~.
\end{split}
\end{align}
For the DA method, $Q^2$ and $x$ can be related to polar angles of the outgoing electron and proton as
\begin{align}
\begin{split}
y_{DA} &= \frac{\tan(\theta_p/2)}{\tan(\theta_e/2) + \tan(\theta_p/2)}~, \\
Q^2_{DA} &= 4E_e^2 \frac{\cot(\theta_e/2)}{ \tan(\theta_e/2) + \tan(\theta_p/2)}~, \\
x_{DA} &= \frac{Q^2_{DA}}{s y_{DA}}~.
\end{split}
\end{align}

\begin{figure}[ht]	
	\centering
	\begin{subfigure}[b]{0.49\textwidth}
	    \centering
		\includegraphics[keepaspectratio=true,width=2.9in,page=10]{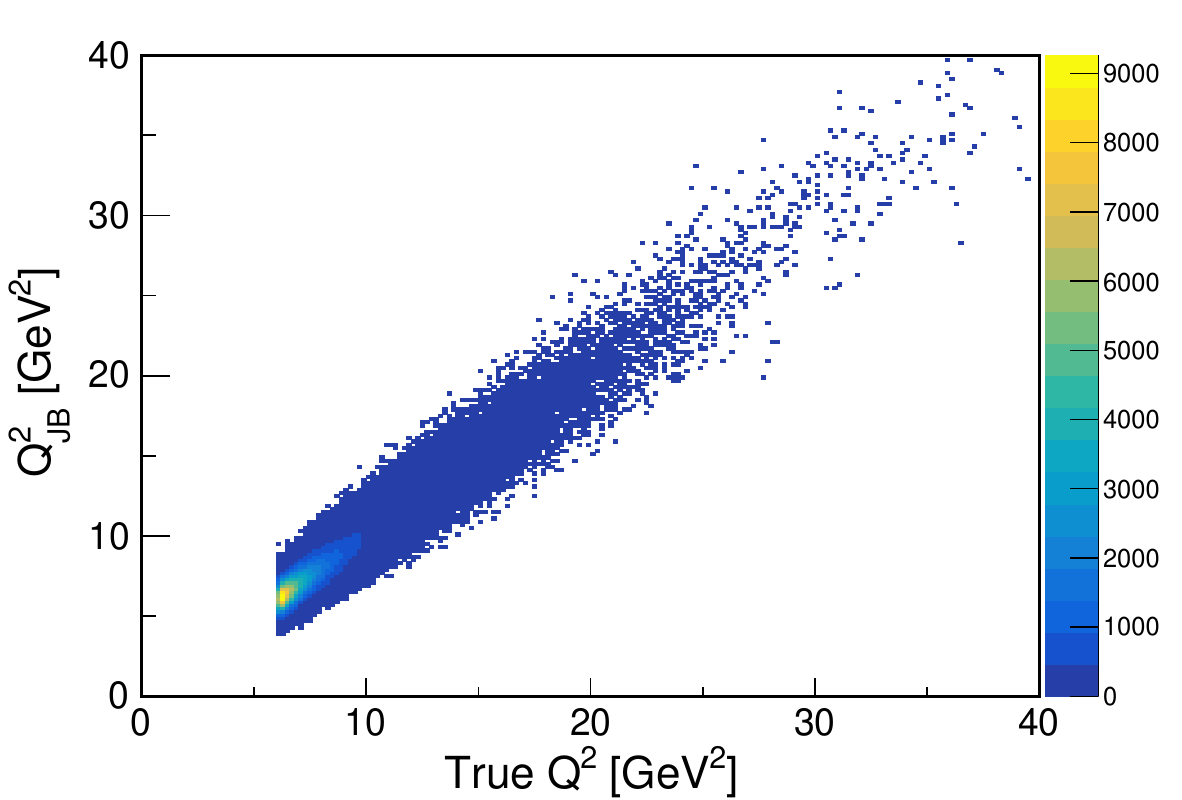}
		\caption{JB method} \label{fig:reconstruction_two_a}
	\end{subfigure}
	\hfill
	\begin{subfigure}[b]{0.49\textwidth}
	    \centering
		\includegraphics[keepaspectratio=true,width=2.9in,page=11]{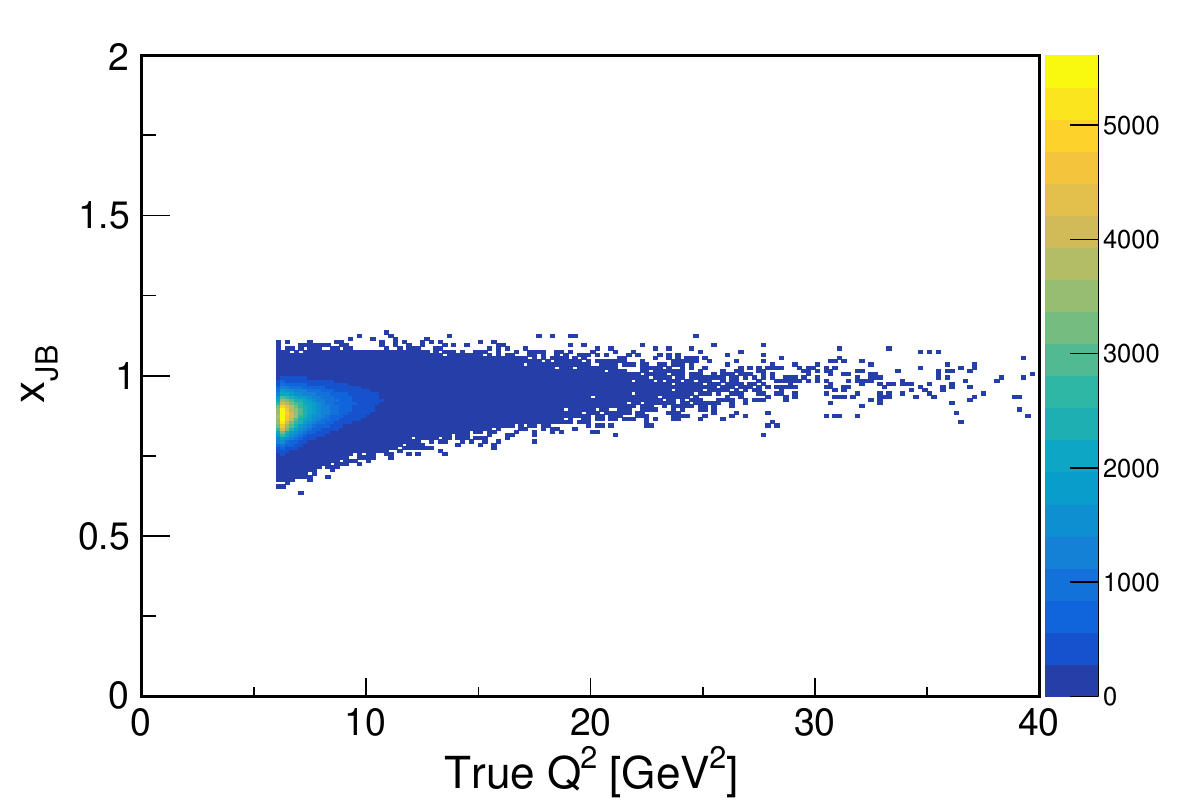}
		\caption{JB method} \label{fig:reconstruction_two_b}
	\end{subfigure}
	\hfill
	\begin{subfigure}[b]{0.49\textwidth}
	    \centering
		\includegraphics[keepaspectratio=true,width=2.9in,page=13]{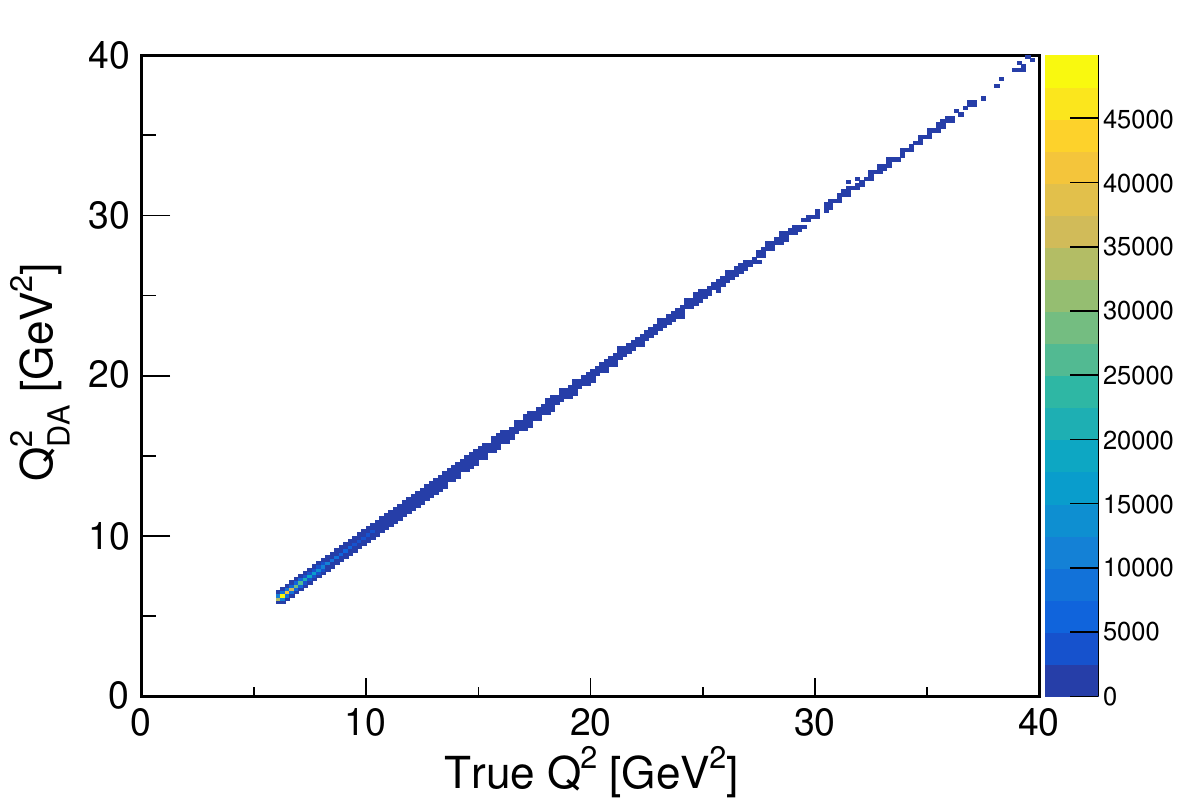}
		\caption{DA method} \label{fig:reconstruction_two_c}
	\end{subfigure}
	\hfill
	\begin{subfigure}[b]{0.49\textwidth}
	    \centering
		\includegraphics[keepaspectratio=true,width=2.9in,page=14]{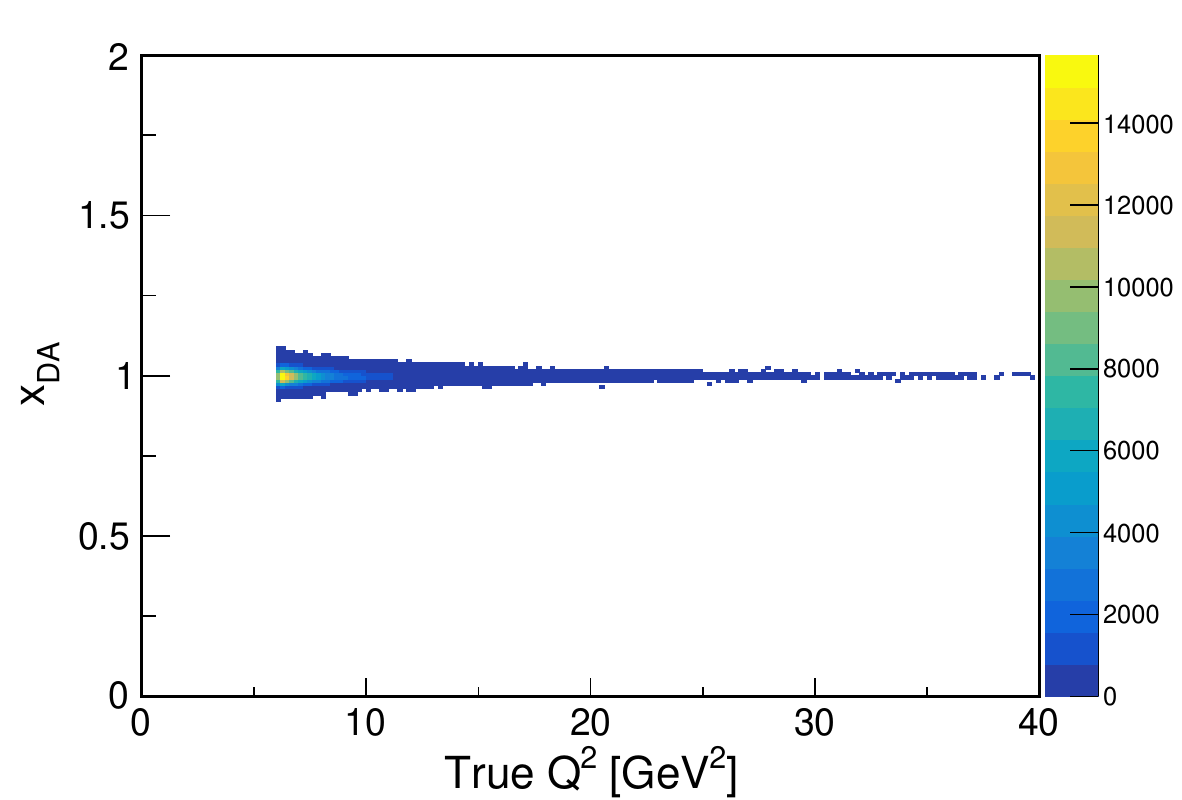}
		\caption{DA method} \label{fig:reconstruction_two_d}
	\end{subfigure}
	\caption{Reconstruction of the kinematic variables relevant to elastic e-p scattering using the JB and DA methods as a function of the true $Q^2$. Top left: $Q^2$ reconstructed using the JB method; top right: $x$ reconstructed using the JB method; bottom left: $Q^2$ reconstructed using the DA method; bottom right: $x$ reconstructed using the DA method. All results are shown for the lowest $\sqrt{s}$ setting $-$ 5 GeV electron on 41 GeV proton $-$ which is the only EIC energy setting where the outgoing elastic proton is within the assumed central detector acceptance. The events are generated using the Born-level elastic generator introduced in section~\ref{sec:kinematics}.} \label{fig:reconstruction_two}
\end{figure}

As both the JB and DA methods require the proton to be reconstructed, we only study the 5 GeV electron on 41 GeV proton energy setting $-$ where the elastically scattered proton is within the assumed central detector acceptance. The results of these two reconstruction methods are shown in figure~\ref{fig:reconstruction_two}. The JB reconstruction of $Q^2$ is worse than the electron method, but the reconstruction of $x$ produces a visible elastic peak. As can be clearly seen, however, the DA method is superior to the elctron and JB approaches for both $Q^2$ and $x$ reconstruction. For the DA method, the resolution $Q^2$ is much better than the binning width chosen in figure~\ref{fig:events_one}; the reconstruction of the elastic peak is very good over the entire range of $Q^2$, and should allow for a precision selection of the elastic peak and tail. 

\clearpage

\section{Background} \label{sec:bg}
In order to accurately measure the e-p elastic cross section for $6~(GeV/c)^2 < Q^2 < 40~(GeV/c)^2$, the elastic events need to be separated from the inelastic background with a very high efficiency. In this section, we consider three questions:

\begin{enumerate}
    \item How well can the elastic events be separated from higher-$Q^2$ DIS events where the scattered electron goes into the central detector acceptance?
    \item Can the high-$Q^2$ elastic events that we want to reconstruct be separated from minimum-bias DIS events?
    \item What are the particle identification (PID) challenges for the elastically scattered electron and proton?
\end{enumerate}

\begin{figure}[ht]	
	\centering
	\begin{subfigure}[b]{0.49\textwidth}
	    \centering
		\includegraphics[keepaspectratio=true,width=2.9in,page=3]{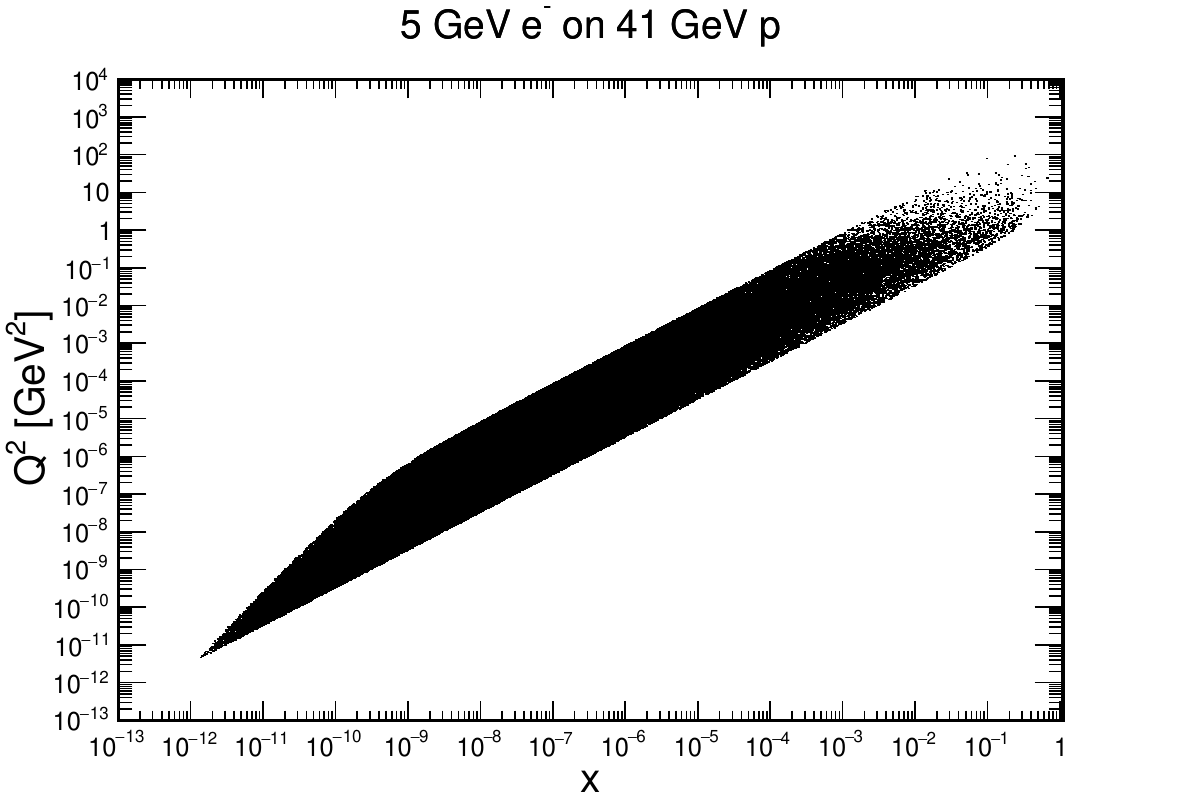}
	\end{subfigure}
	\hfill
	\begin{subfigure}[b]{0.49\textwidth}
	    \centering
		\includegraphics[keepaspectratio=true,width=2.9in,page=4]{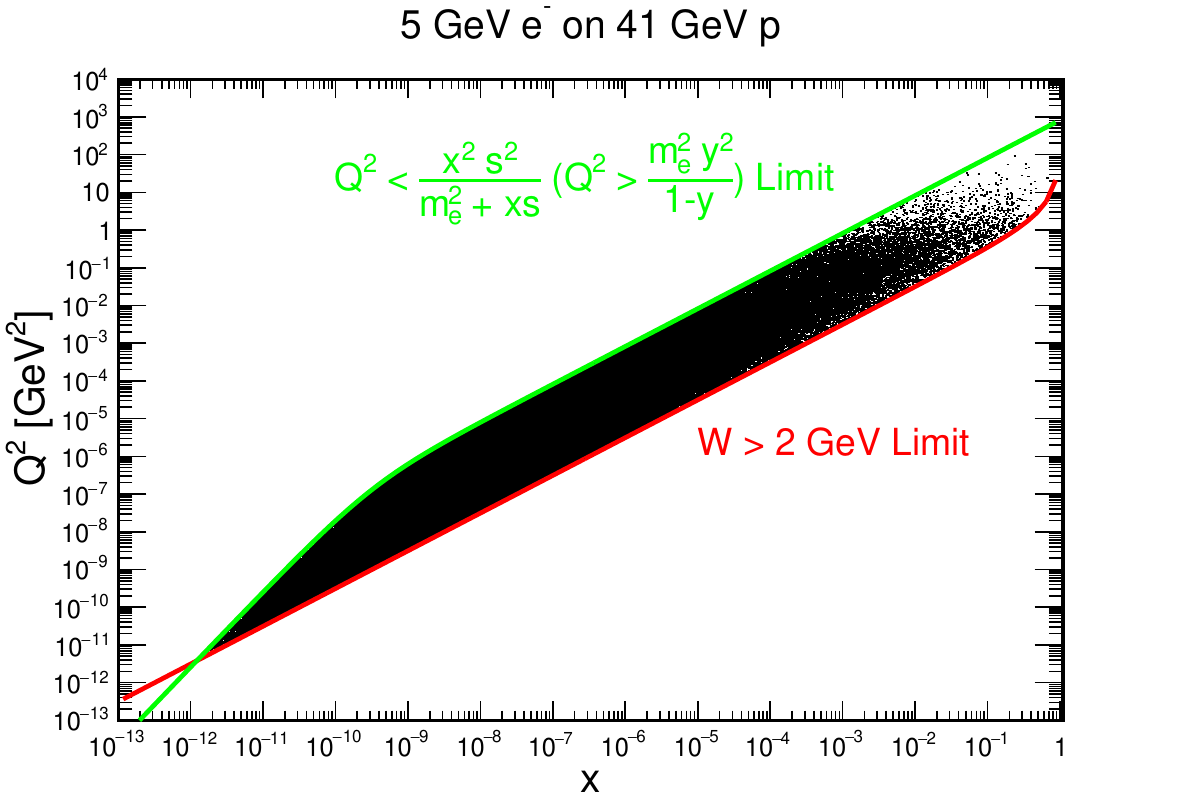}
	\end{subfigure}
	\caption{Kinematic converge in the $Q^2-x$ plane for minimum-bias events for the 5 GeV electron on 41 GeV proton energy setting. The events were generated using the \textit{Pythia6} event generator and show two kinematic limits. These kinematic limits are the requirement that the invariant mass of the final hadronic state is greater than 2~$GeV/c^2$ $-$ the red curve in the right panel $-$ which is a limit imposed by \textit{Pythia6}, and the photoproduction minimum requirement that $Q^2>\frac{m_e^2y^2}{1-y}$ (equivalent to $Q^2 < \frac{x^2s^2}{m_e^2+xs}$) $-$ the green curve in the right panel.} \label{fig:bg_one}
\end{figure}

To answer these questions, we generate several DIS samples using the \textit{Pythia6} event generator, and then pass these events through the detector simulation described in section~\ref{sec:detector}. We only perform these studies for the 5~GeV electron on 41~GeV proton energy set, for the reasons discussed in section~\ref{sec:reconstruction}. For question 1, we generate two sets of data consisting of 100 million events each $-$ one with $Q^2 > 0.5~(GeV/c)^2$, and the other with $Q^2 > 3.0~(GeV/c)^2$. For question 2, we generate 500 million minimum-bias events. The kinematic coverage in the $Q^2-x$ space of these minimum-bias events is shown is figure~\ref{fig:bg_one}. As can be seen, the generated data is bounded by two limits. The red curve shows the requirement final hadronic state mass to be greater than 2~$GeV/c^2$. This appears to be a hard cutoff imposed by \textit{Pythia6} on the all events it generates, and the upshot is that we are not studying potential background in the resonance region. The green curve shows the photoproduction limit for electron-proton scattering, which is a function of the electron mass and the inelasticity variable. 

\begin{figure}[ht]	
	\centering
	\begin{subfigure}[b]{0.49\textwidth}
	    \centering
		\includegraphics[keepaspectratio=true,width=2.9in,page=2]{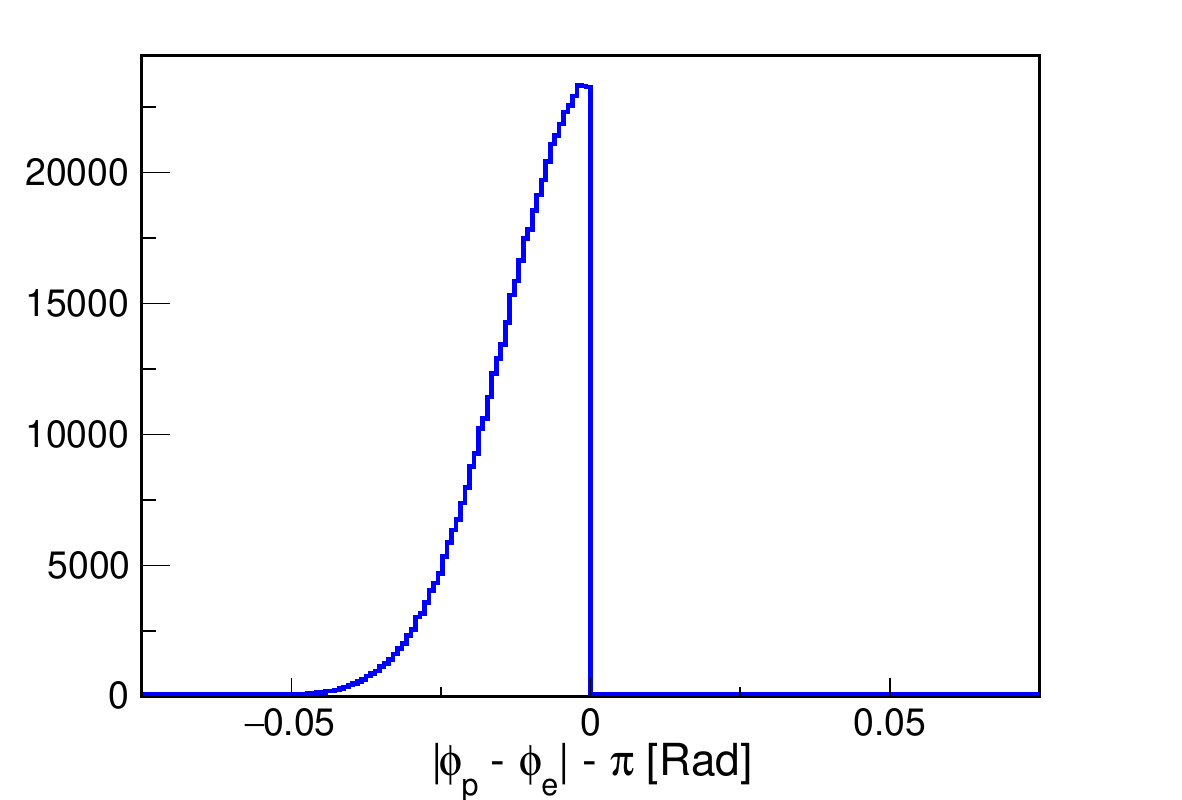}
	\end{subfigure}
	\hfill
	\begin{subfigure}[b]{0.49\textwidth}
	    \centering
		\includegraphics[keepaspectratio=true,width=2.9in,page=3]{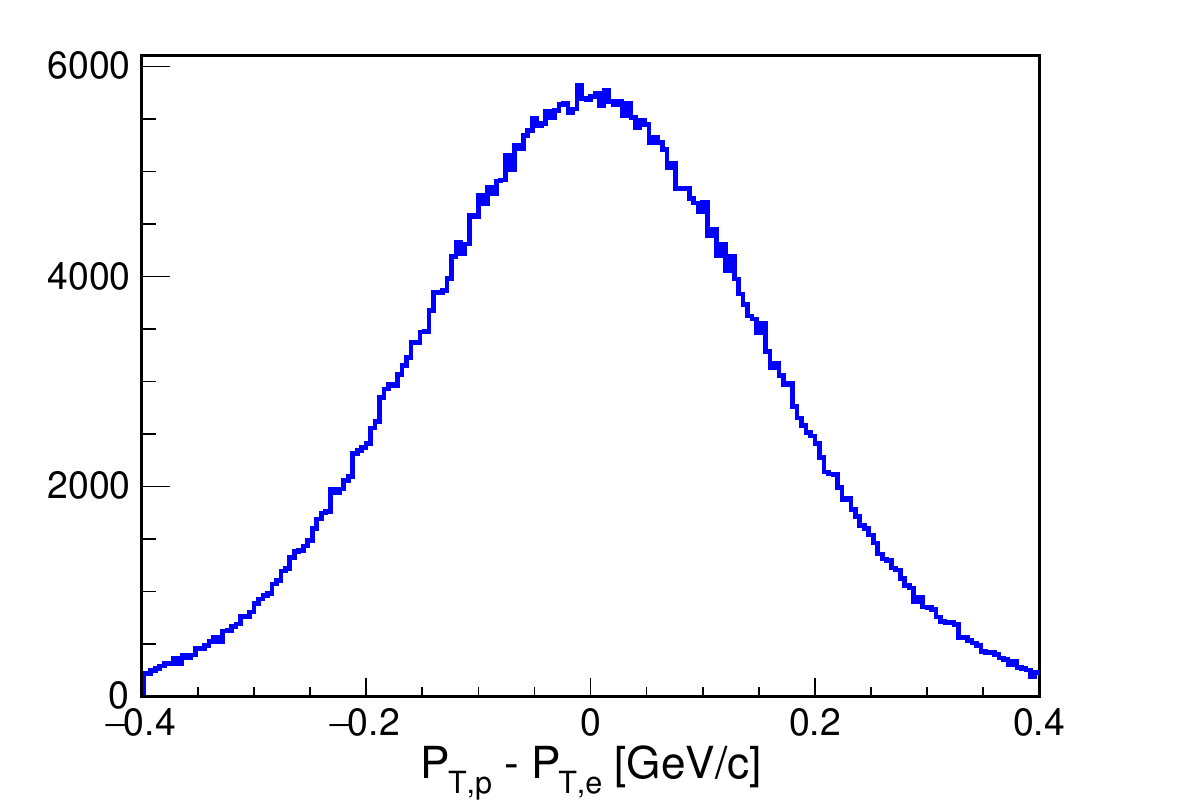}
	\end{subfigure}
	\hfill
	\begin{subfigure}[b]{0.49\textwidth}
	    \centering
		\includegraphics[keepaspectratio=true,width=2.9in,page=4]{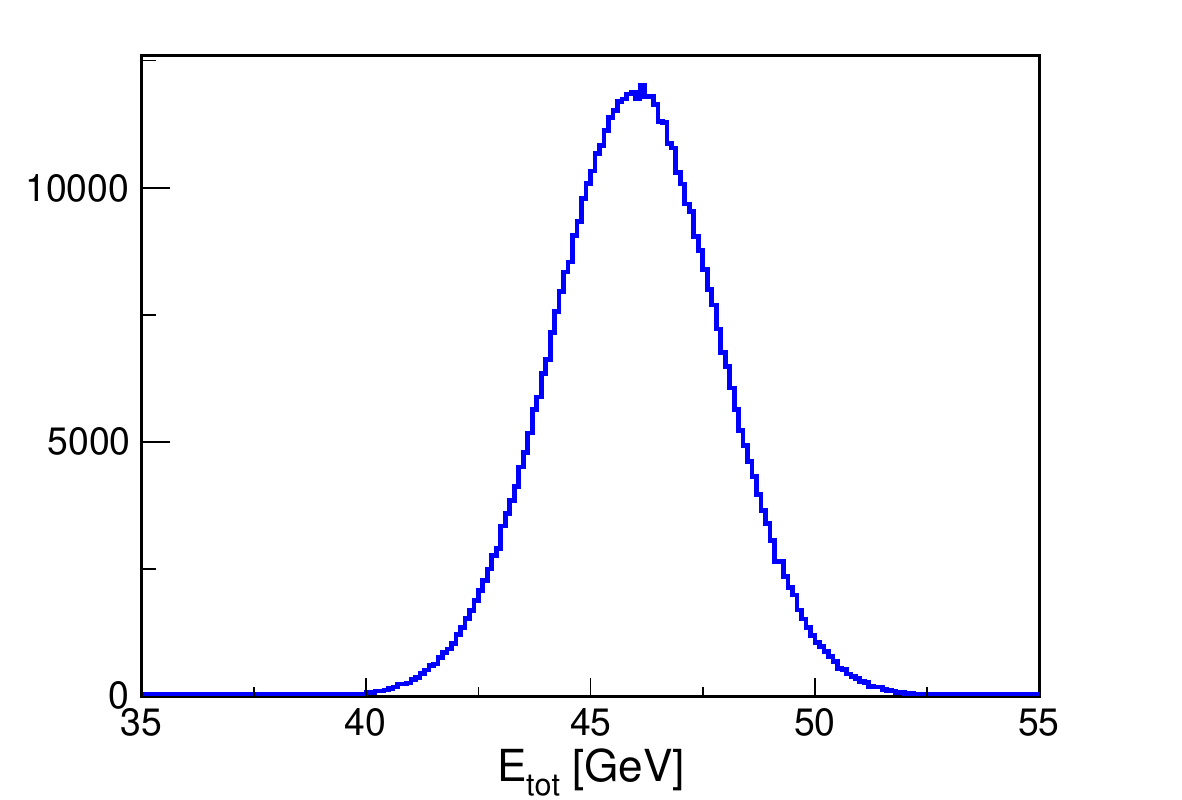}
	\end{subfigure}
	\hfill
	\begin{subfigure}[b]{0.49\textwidth}
	    \centering
		\includegraphics[keepaspectratio=true,width=2.9in,page=5]{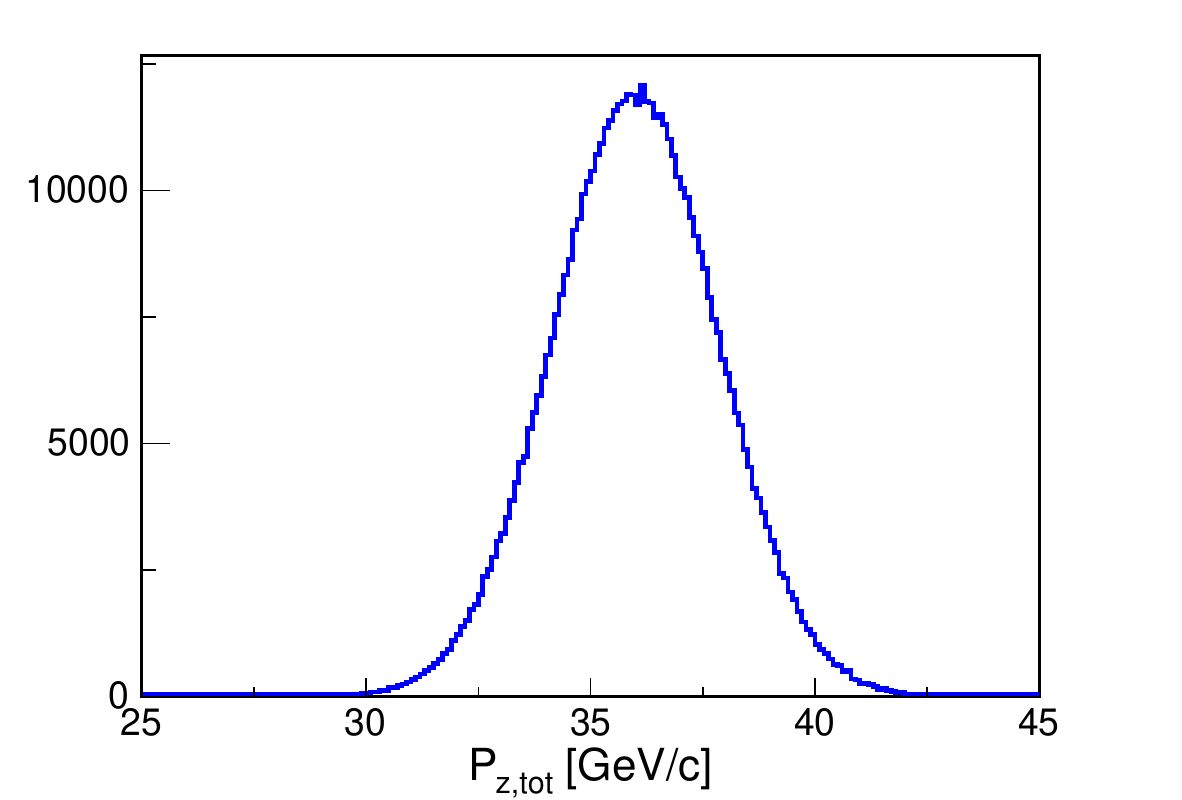}
	\end{subfigure}
	\caption{Reconstructed elastic event topology for the 5 GeV electron on 41 GeV proton energy setting. Top left: difference between the azimuthal angle of the outgoing elastic electron and proton; top right: difference between the electron and proton trasverse momentum; bottom left: total energy of the electron and proton; bottom right: total longitudinal momentum of the electron and proton. All events use the elastic generator discussed in section~\ref{sec:kinematics}, with the events passed through the fast detector simulation package described in section~\ref{sec:detector}. All four plots require both the electron and proton to be reconstructed within the assumed central detector acceptance, which is satisfied for approximately $Q^2 > 6~(GeV/c)^2$.} \label{fig:bg_two}
\end{figure}

We first consider the higher-$Q^2$ DIS background, where the scattered electron is in the central detector. We assume that electrons and protons are correctly identified for both the elastic events and the DIS events. For DIS events where multiple electrons are reconstructed within the central detector acceptance, we assume that the one corresponding to the scattered electron has been correctly identified. We impose a series of selection cuts (vetos) which the elastic events survive with a very large efficiency, and then check how well these cuts remove the inelastic background. The following selection criteria are applied:

\begin{enumerate}
    \item For a given event, only the scattered electron and a single proton are reconstructed in the central detector.
    \item The electron and proton are co-planar. The difference between the reconstructed azimuthal angles of the electron and proton for the elastic events is shown in the top left panel of figure~\ref{fig:bg_two}. The selection criterion applied is $|\phi_p - \phi_e| - \pi > -0.05~Rad$.
    \item The electron and proton transverse momenta balance. The difference between the reconstructed transverse momenta of the electron and proton for the elastic events is shown in the top right panel of figure~\ref{fig:bg_two}. The selection criterion applied is $|P_{T,p} - P_{T_e}| > 0.05~GeV/c$.
    \item The total energy of the detected electron and proton equals the total energy of the incoming beams. The total reconstructed energy of the outgoing elastic electron and proton is shown in the bottom left panel of figure~\ref{fig:bg_two}. As can be seen, the reconstructed energy is centered at 46 GeV, as it should be for a 5 GeV electron on 41 GeV proton beam. The selection criterion applied is $38~GeV < E_{tot} < 54~GeV$.
    \item The total longitudinal momentum ($P_z$) of the detected electron and proton equals the total momentum of the incoming beams. The total reconstructed $P_z$ of the outgoing elastic electron and proton is shown in the bottom right panel of figure~\ref{fig:bg_two}. The reconstructed longitudinal momentum is centered at 36 GeV/c. The selection criterion applied is $30~GeV < P_{z,tot} < 42~GeV$.
\end{enumerate}

The results of applying these selection cuts in succession to the higher-$Q^2$ DIS events are shown in table~\ref{tab:bg_1}. As can be seen in the table, the cuts are successful in vetoing all 100 million inelastic events for both the $Q^2 > 0.5~(GeV/c)^2$ and the $Q^2 > 3.0~(GeV/c)^2$ data sets. 100 million events corresponds to an integrated luminosity of 0.14~fb$^{-1}$ for $Q^2 > 0.5~(GeV/c)^2$ and 0.96~fb$^{-1}$ for $Q^2 > 3.0~(GeV/c)^2$. The expected event counts scaled to $100 fb^{-1}$ after each cut are also shown in table~\ref{tab:bg_1}.

\begin{table}[ht]
\centering
\begin{tabular}
{c||c||c}
Veto & Events with $Q^2 > 0.5~GeV^2$ & Events with $Q^2 > 3.0~GeV^2$ \\
& (Scaled to $100~fb^{-1}$) &(Scaled to $100~fb^{-1}$) \\
\hline
None & $100 \times 10^{6} \left(7.16 \times 10^{10} \right)$ & $100 \times 10^{6} \left(1 \times 10^{10} \right)$\\
1 & $306060 \left(2.19 \times 10^{8}\right)$ & $213440 \left(2.22 \times 10^{7}\right)$\\
1+2& 32 (22912)& 0 (0) \\
1+2+3 &13 (9308) & 0 (0)  \\
1+2+3+4 &0 (0) &0 (0) \\
1+2+3+4+5 & 0 (0) &0 (0)
\end{tabular} 
\caption{Effect of the selection criteria (vetos) described in the text on suppressing the higher-$Q^2$ inclusive DIS event background. Column 1 shows the vetos applied; column 2 and 3 show the number of events passing the specific set of vetos, for the $Q^2 > 0.5~(GeV/c)^2$ and the $Q^2 > 3.0~(GeV/c)^2$ simulation data sets, respectively.}
\label{tab:bg_1}
\end{table}

For the minimum-bias DIS background shown in figure~\ref{fig:bg_one}, we generated 500 million \textit{Pythia6} events, which corresponds to an integrated luminosity of only 0.006~fb$^{-1}$. The large minimum-bias cross section is dominated by low-$Q^2$ events where the scattered electron is not in the central detector acceptance. The background to the high-$Q^2$ elastic events therefore comes from electrons in the central detector acceptance which are produced in electroweak processes. The question we consider is whether these 'decay' electron background events can be removed as efficiently as the high-$Q^2$ DIS background events using the same selection criteria.

\begin{figure}[ht]
\centering
\includegraphics[width=0.75\textwidth,page=2]{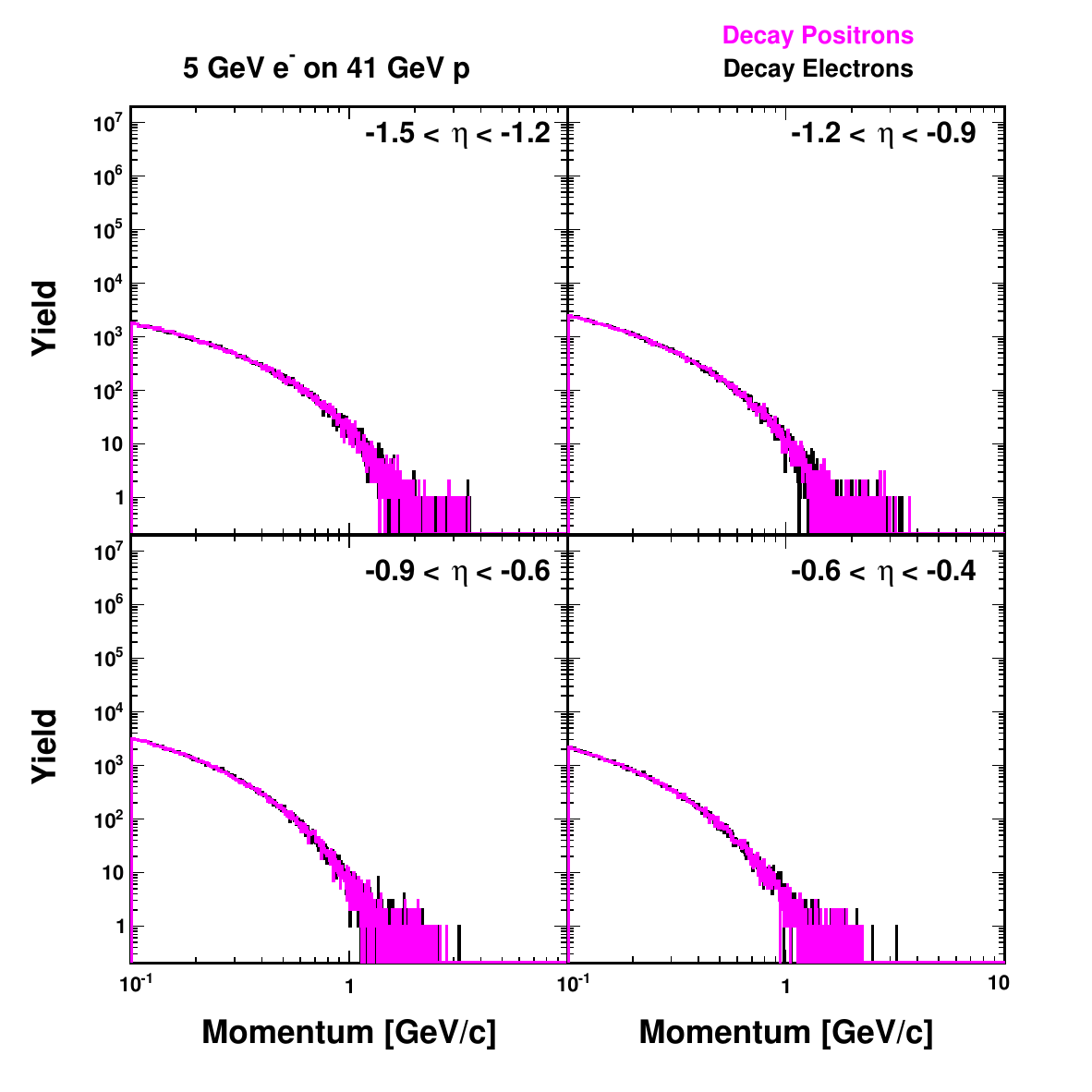}
\caption{Spectra of electrons other than the scattered electron and all positrons for minimum-bias DIS events. The spectra are shown in four different angular bins within the central detector acceptance which are relevant for the high-$Q^2$ elastic electron.}
\label{fig:bg_three}
\end{figure}

One experimental observation is that these 'decay' electrons are produced at the same rate as positrons in minimum-bias electron-proton DIS events. This is shown for different central detector angular ranges in figure~\ref{fig:bg_three} for the \textit{Pythia6} minimum-bias events. In the experiment, the surviving background from electrons other than the scattered electron can be determined by studying the positron yield. In particular, we apply the same selection criteria listed above, replacing the scattered electron in the analysis with a positron. For example, the first selection criterion is now requires the event to have a single positron and a single proton in the central detector acceptance. Table~\ref{tab:bg_2} shows the results of applying these cuts in succession to the minimum-bias sample. The cuts successfully veto all 500 million minimum-bias DIS background events.

\begin{table}[ht]
\centering
\begin{tabular}
{c||c}
Veto & Total Events (Scaled to $100~fb^{-1}$)  \\
\hline
None & $500 \times 10^{6} \left(8 \times 10^{12}\right)$ \\
1 & $148 \left(2.37 \times 10^{6}\right)$ \\
1+2 &1 (16000) \\
1+2+3 &1  (16000)  \\
1+2+3+4 & 0 (0)\\
1+2+3+4+5 & 0 (0) 
\end{tabular} 
\caption{Effect of the selection criteria (vetos) described in the text on suppressing the minimum DIS event background. Column 1 shows the vetos applied; column 2 shows the number of events passing the specific set of vetos.}
\label{tab:bg_2}
\end{table}

\begin{figure}[ht]
\centering
\includegraphics[width=0.75\textwidth,page=1]{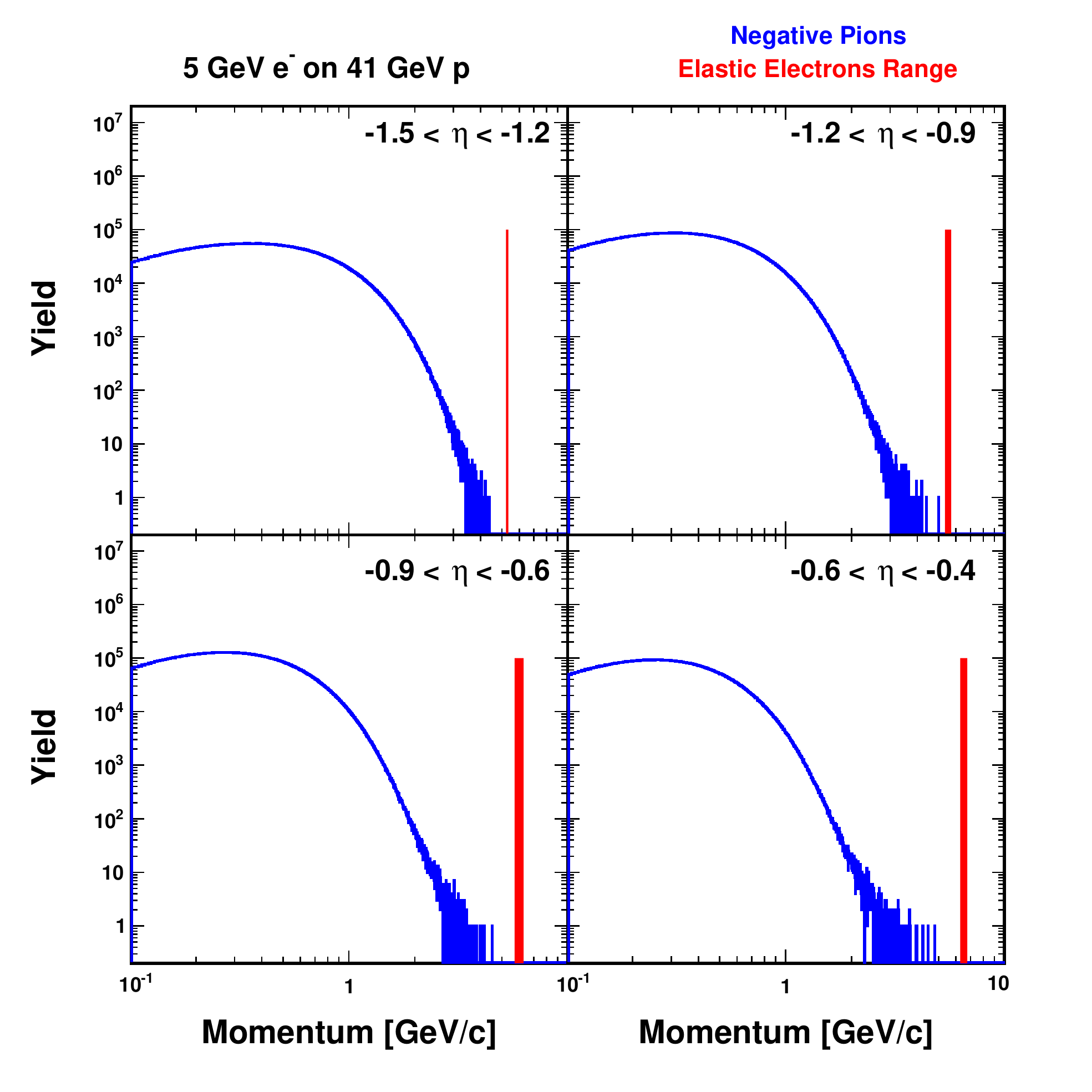}
\caption{Raw spectra of negatively charged pions generated using the mimimum-bias \textit{Pythia6} simulation. The pion spectrum is shown in four different angular bins within the central detector acceptance. The momentum range of the elastic electrons for each angular bin is also shown. The pion spectra are shown for 500 million minimum-bias events. If $100~fb^{-1}$ of data is collected, these spectra will be scaled up by a factor of $10^{4}$. Detector-based (e.g. calorimeter and Cherenkov) electron PID methods should be able suppress the raw pion background by a factor of $10^{3} - 10^{4}$, before any additional elastic selection cuts are applied.}
\label{fig:bg_four}
\end{figure}

The final question we consider is the PID requirements for the high-$Q^2$ elastic electron and proton in the central detector. For the outgoing elastic electron, the primary PID challenge comes from the negative pions produced in the minimum-bias DIS events. Figure~\ref{fig:bg_four} shows this raw pion spectra as a function of momentum for different angular ranges in the central detector acceptance using the 500 million generated minimum-bias events. The momentum range of the elastic electrons is also shown for each angular bin. The elastic electrons will have larger momentum than the raw negative pion background, and therefore removing the pion background from the elastic sample should be straightforward.

\begin{figure}[ht]
\centering
\includegraphics[width=0.75\textwidth,page=3]{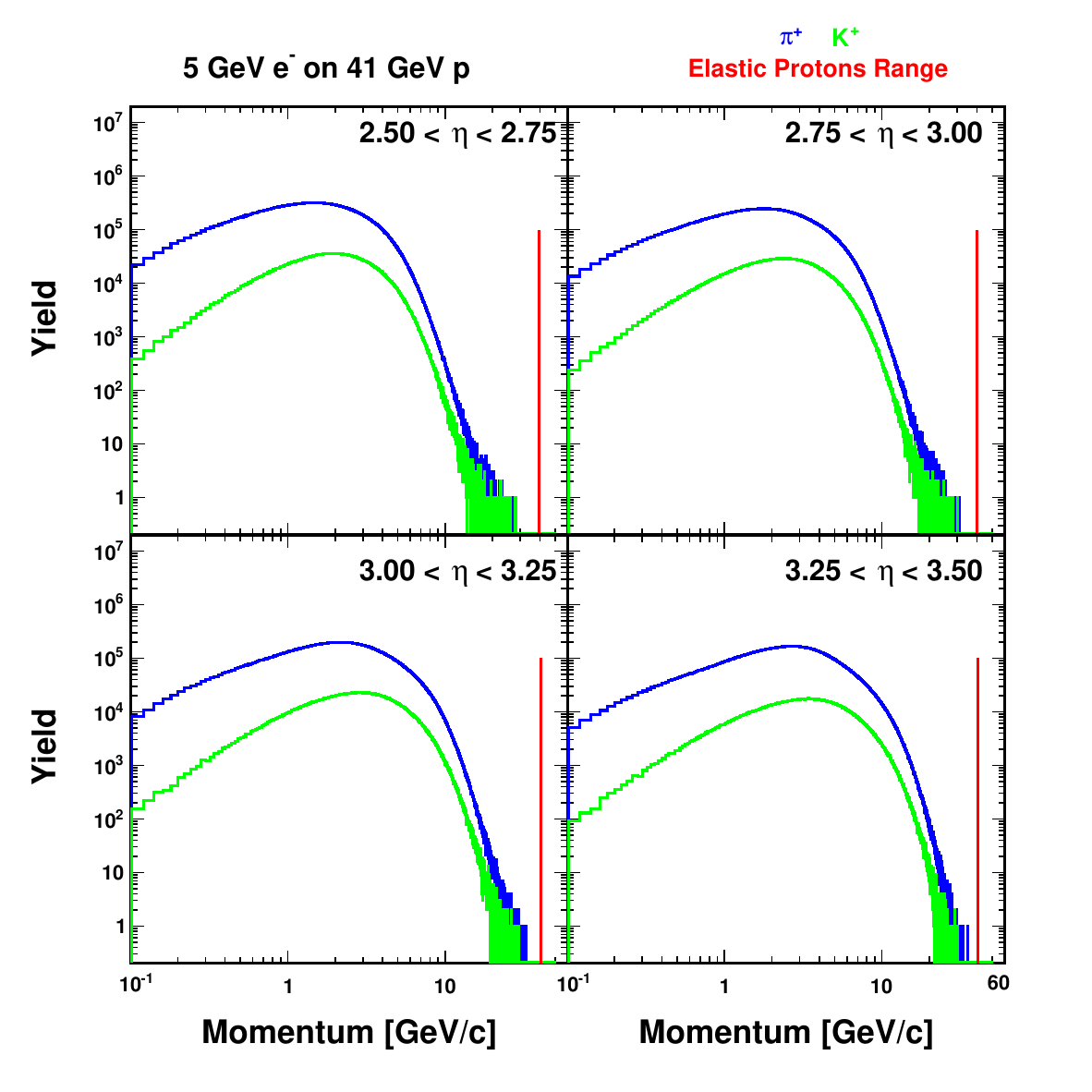}
\caption{Raw spectra of positively charged pions (blue curves) and kaons (green curves) generated using the mimimum-bias \textit{Pythia6} simulation. The spectra are shown in four different angular bins within the central detector acceptance. The momentum range of the elastic protons for each angular bin is also shown (red region). The pion spectra are shown for 500 million minimum-bias events. If $100~fb^{-1}$ of data is collected, these spectra will be scaled up by a factor of $10^{4}$.}
\label{fig:bg_five}
\end{figure}

As for the outgoing elastic proton, the challenge is separating the proton from positively charged pions and kaons. As the proton will have a momentum of about 40~$GeV/c$ and be detected in the central detector after a flight path of a few meters, using time-of-flight techniques would require sub-picosecond resolution is not practical. Separation of the elastic proton from pions and kaons would be best archived using a gaseous Ring Imaging Cherenkov detector~\cite{Akopov:2000qi}. Similar to the case of the elastic electron, the background particles for the elastic proton tend to be at lower momentum, as shown in figure~\ref{fig:bg_five}. So the requirement of sufficiently large momentum will be sufficient to remove the vast majority of the charged pion and kaon background.

\clearpage

\section{Conclusion} \label{sec:concl}
In this work, we have considered the potential of the EIC to make high-$Q^2$ elastic electron-proton unpolarized cross section measurements. By considering the elastic event yields and kinematics, as well as questions of reconstruction, we found that the EIC has the capability to measure the elastic cross section in the momentum transfer range of $6~(GeV/c)^2 < Q^2 < 40~(GeV/c)^2$ at $\epsilon \sim 1$.

As the EIC project proceeds, additional work on this topic would be useful. In particular, this study did not incorporate QED radiative effects for the elastic events~\cite{Gramolin:2014pva} or for the background events. In addition, background from the resonance region was not considered. Future reconstruction and background studies should take these effects into account.

This work focused on the case when both the elastically scattered electron and proton are in the central detector acceptance, which restricted the study to the lowest EIC beam energy setting. As the luminosity of the EIC is expected to increase some at the higher $\sqrt{s}$ settings, using these higher energies for the elastic cross-section measurements will be necessary for good statistical precision at high $Q^2$. Once a near-final design for an EIC detector including the far-forward region is complete, these elastic scattering studies should be repeated at higher energies using the full detector simulation. Studies of elastic scattering at future lower-energy electron-proton colliders~\cite{Anderle:2021wcy} can also be performed.

This work also did not perform impact studies on the elastic form factors using the generated pseudo-data. This would have required some realistic consideration of the possible systematic uncertainties on the cross-section measurements.

\appendix
\section{Effects of non-zero crossing angle} \label{sec:app_ca}
The work presented in this paper was performed assuming co-linear electron and proton beams. In reality, the EIC will have a non-zero crossing angle. The overall effects of the non-zero crossing angle and other beam parameters were studied in Ref.~\cite{EIC_angle}. Here we consider only how a non-zero crossing angle effects the event rate and kinematics of the elastic electron and proton.

In the EIC detector frame, assume the incoming proton beam has a 4-momentum ($p_x$,$p_y$,$p_z$,$E$) of $p_i = E_p (\sin\theta_c,0,\cos\theta_c,1)$ and the incoming electron has a 4-momentum of $e_i = E_e (0,0,-1,1)$. $\theta_c$ is the crossing angle, which will be about 25~mRad at the first interaction point of the EIC.

A linear combination of the beam 4-vectors $p_i$ and $e_i$ can be written as 
\begin{equation}
b = c_0 \times p_i + c_1 \times e_i ~,
\end{equation}
where $c_0$ and $c_1$ are constants. With a Lorentz transform into the rest frame of $b$, the beams $p_i$ and $e_i$ will be co-linear. In order to assess the effect of the crossing angle on the results shown in this paper, we first calculate the values of $c_0$ and $c_1$ that will give the minimum change in the energies of the two beams between the two reference frames. This will happen when the Lorentz Boost is a zero at first-order in the z direction. The velocity of this new frame with respect to the lab frame is
\begin{equation} \label{boost_eqn}
v_{b} / c = = \left(\frac{\sin\theta_{c}}{2},0,\frac{\cos\theta_{c} -1}{2}\right) \approx (\theta_{c}/2,0,-\theta_{c}^{2}/4) ~ ,
\end{equation}
where $c$ is the speed of light.

The energies of the electron and proton beams in this co-linear frame differ from their respective energies in the lab frame by $\mathcal{O}(\theta_c^2)$, which, for a crossing angle of 25~mRad, is at the level of $10^{-4}$. Therefore, the error on the cross section shown in figure~\ref{fig:events_one} due to generating the events with the lab frame energies instead of the co-linear frame energies is negligible.

We next consider the difference in the kinematics of the outgoing elastic electron and proton between the lab frame and the same minimally transformed co-linear frame $b$. Since the beam energies are nearly equivalent in both frames, the kinematic distributions of the outgoing electron and proton shown in figures~\ref{fig:events_two} and \ref{fig:events_three} are accurate in the co-linear frame. In order relate the kinematics of a particle in $b$ to its kinematics in the lab frame, we use the Lorentz transformation given in Eq.~\eqref{boost_eqn} and then apply a rotation so that the proton beam is along the positive z axis in frame $b$. The 4-momentum of a particle in frame $b$ is then related to its 4-momentum in the lab frame by
\begin{equation}
\begin{bmatrix}
E \\
p_x \\
p_y \\
p_z
\end{bmatrix}_{b}
= L_{b+r}
\begin{bmatrix}
E \\
p_x \\
p_y \\
p_z
\end{bmatrix}_{lab} ~ ,
\end{equation}
where $L_{b+r}$ is the first-order transformation matrix and is given by
\begin{equation}
L_{b+r} = 
\begin{bmatrix}
1 & -\theta_c/2 & 0 & 0 \\
-\theta_c/2 & 1 & 0 & -\theta_c/2 \\
0 & 0 & 1 & 0 \\
0 & \theta_c/2 & 0 & 1
\end{bmatrix} ~.
\end{equation}

The transformation from the co-linear frame $b$ to the lab frame is given by the inverse of $L_{b+r}$, which, to first-order in $\theta_c$ is given by
\begin{equation}
L^{-1}_{b+r} = 
\begin{bmatrix}
1 & \theta_c/2 & 0 & 0 \\
\theta_c/2 & 1 & 0 & \theta_c/2 \\
0 & 0 & 1 & 0 \\
0 & -\theta_c/2 & 0 & 1
\end{bmatrix} ~.
\end{equation}

The energy and angles of the particle in the lab frame are the relevant quantities for questions relating to detector acceptance and resolution. Therefore, it is important to determine how much the kinematics differ between lab and the generated quantities shown in figures~\ref{fig:events_two} and \ref{fig:events_three}. For ultra-relativistic particles (i.e. the particle's mass is much less than its energy) such as the elastically scattered electron and proton, the 4-momentum in frame $b$ can be written as
\begin{equation}
\begin{bmatrix}
E \\
p_x \\
p_y \\
p_z
\end{bmatrix}_{b}
=
E_{gen}
\begin{bmatrix}
1 \\
\sin\theta_{gen} \cos\phi_{gen} \\
\sin\theta_{gen} \sin\phi_{gen}\\
\cos\theta_{gen}
\end{bmatrix}~,
\end{equation}
where $E_{gen}$, $\theta_{gen}$, and $\phi_{gen}$ are the energy, polar angle, and azimuthal angle of the particle in co-linear frame $b$, respectively. The 4-momentum of the particle in the lab frame can then be calculated using the transformation matrix $L^{-1}_{b+r}$, and expressed as
\begin{equation}\label{frame_relation}
\begin{bmatrix}
E \\
p_x \\
p_y \\
p_z
\end{bmatrix}_{lab}
=
E_{gen}
\begin{bmatrix}
1 + \left(\theta_c/2\right) \sin\theta_{gen} \cos\phi_{gen} \\
\theta_c/2 + \sin\theta_{gen} \cos\phi_{gen} + \left(\theta_c/2\right) \cos\theta_{gen}  \\
\sin\theta_{gen} \sin\phi_{gen} \\
-\left(\theta_c/2\right) \sin\theta_{gen} \cos\phi_{gen} + \cos\theta_{gen}
\end{bmatrix}~.
\end{equation}

The largest difference in the particle's energy between frame $b$ and the lab frame will occur when $ \phi_{gen} = 0^{o}$ or $180^{o}$ and $\theta_{gen} = 90^{o}$ ($\eta_{gen} = 0$):
\begin{equation}
\left( E_{lab} / E_{gen}\right)_{max} = 1 \pm \theta_c/2 ~.
\end{equation}
For a crossing angle of 25~mRad, the maximum difference in the energy of an high-energy particle between the two frames is 1.25\%.

\begin{figure}[ht]	
	\centering
	\begin{subfigure}[b]{0.49\textwidth}
	    \centering
		\includegraphics[keepaspectratio=true,width=2.9in,page=2]{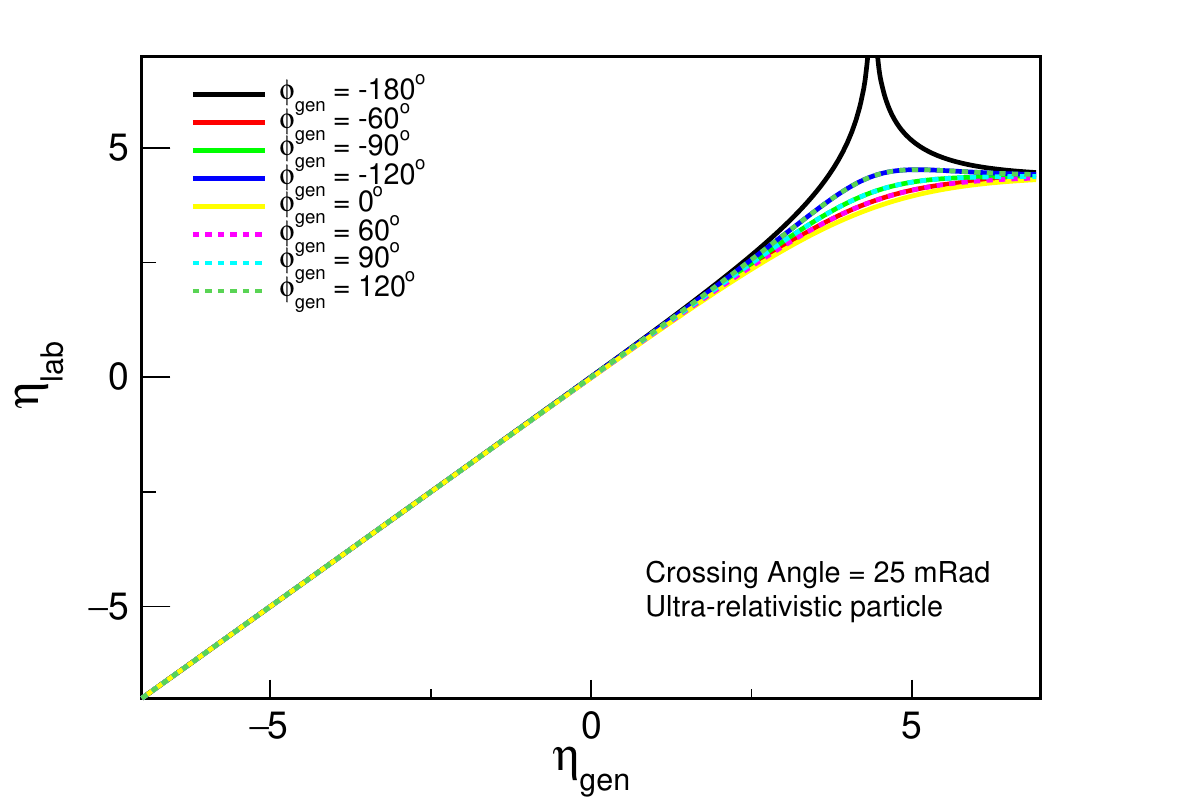}
	\end{subfigure}
	\hfill
	\begin{subfigure}[b]{0.49\textwidth}
	    \centering
		\includegraphics[keepaspectratio=true,width=2.9in,page=3]{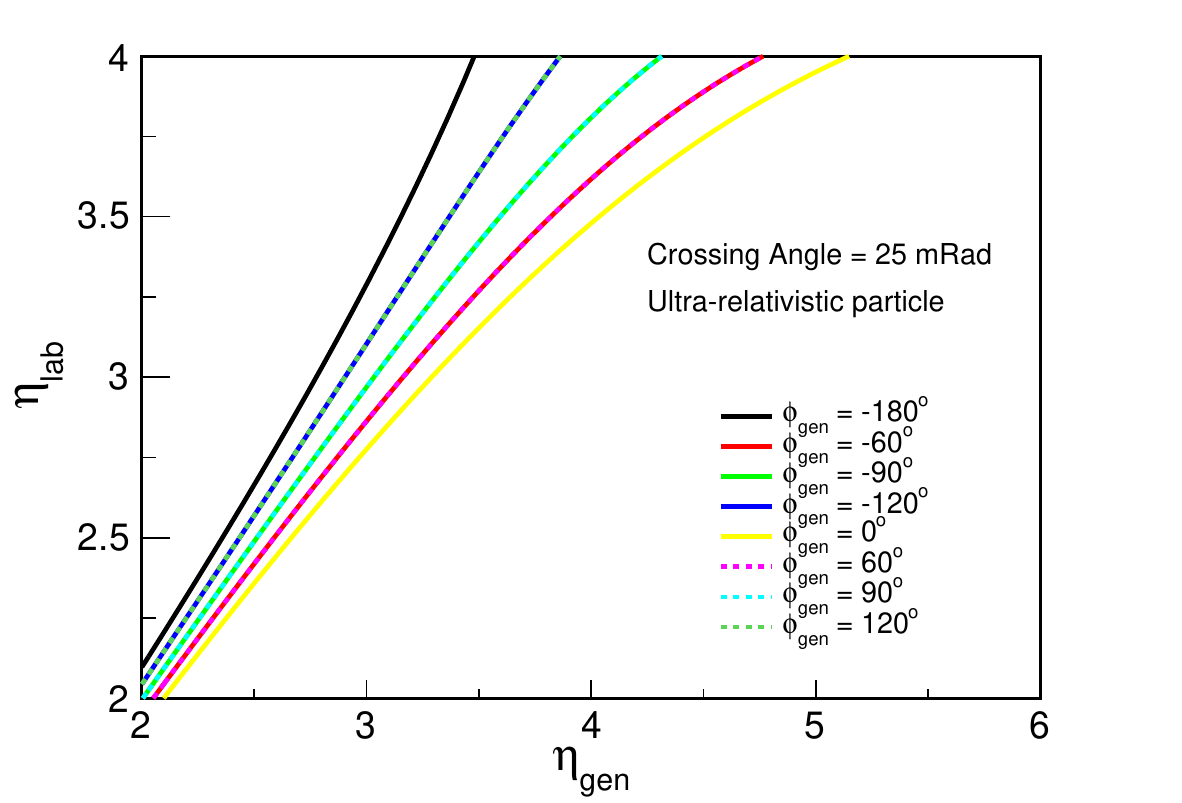}
	\end{subfigure}
	\caption{Dependence of the lab-frame pseudo-rapidity $\eta_{lab}$ on the pseudo-rapidity in the co-linear frame $b$. $\eta_{lab}$ is calculated with the positive z axis anti-parallel to the electron beam direction. The right plot focuses on the positive pseudo-rapidity region, where the dependence on $\phi_{gen}$ is large.} \label{fig:app_one}
\end{figure}

The angles of the particle in the lab frame, $\theta_{lab}$ ($\eta_{lab}$) and $\phi_{lab}$, are functions of the angles in the co-linear frame $b$. In the lab frame, as discussed above, the incoming electron beam will be along the negative z direction and the proton beam will have a small angle ($\theta_c$) with respect to the positive z axis. In figure~\ref{fig:app_one}, $\eta_{lab}$, is calculated as a function of $\eta_{gen}$ for several values of $\phi_{gen}$ using Eq. \eqref{frame_relation}. $\eta_{lab}$ is approximately equivalent to $\eta_{gen}$ for $\eta_{gen} < -1$, but gradually develops a dependence on $\phi_{gen}$ for more positive values of pseudo-rapidity. Since the detector acceptance is set in terms of $\eta_{lab}$ $-$ in this work, for example, we assumed an EIC central detector acceptance of $-3.5 < \eta_{lab} < 3.5$ $-$ the acceptance in $\eta_{gen}$ will have a significant dependence on the azimuthal angle. For elastic e-p scattering, this would manifest itself as a dependence on proton acceptance (and hence the $Q^2$ acceptance) on $\phi_{gen}$.  

\begin{figure}[ht]
\centering
\includegraphics[width=0.75\textwidth,page=4]{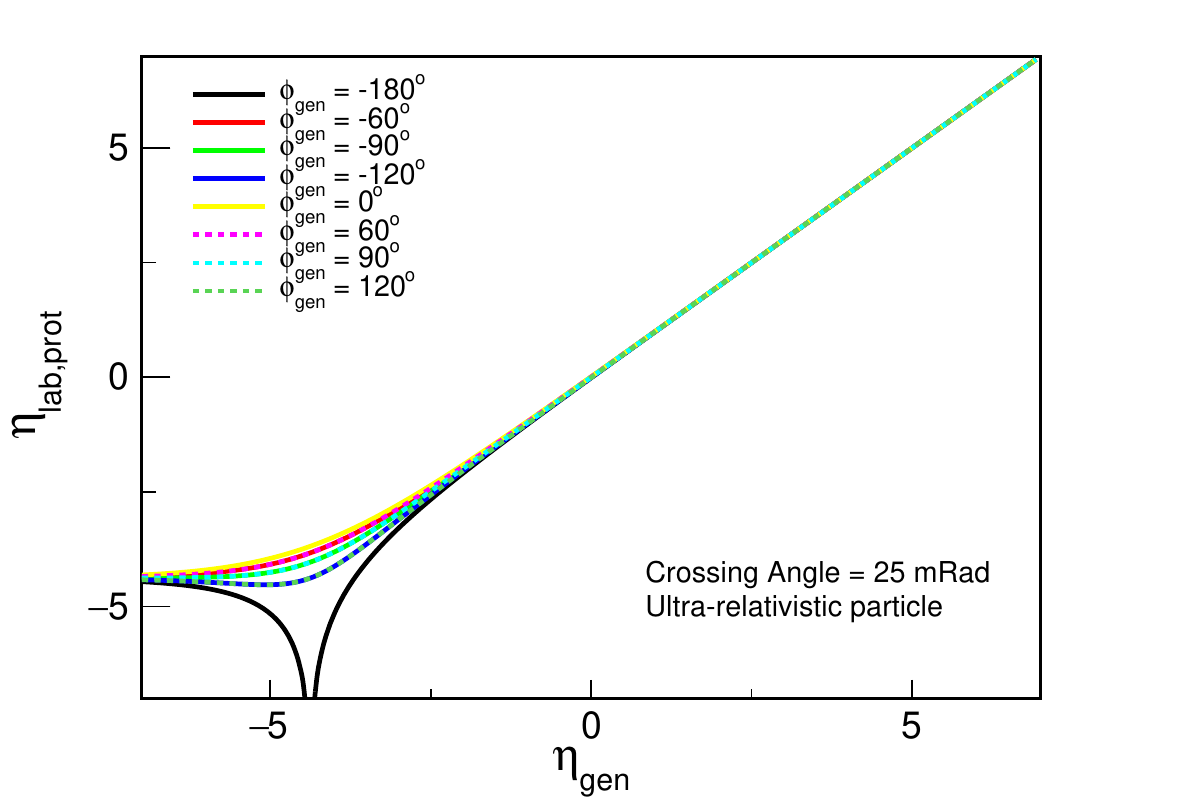}
\caption{Dependence of the lab-frame pseudo-rapidity $\eta_{lab,prot}$ on the pseudo-rapidity in the co-linear frame $b$. Here, the lab-frame pseudo-rapidity is calculated with the positive z axis parallel to the proton beam direction. The figure is clearly a mirror image of the left plot in figure~\ref{fig:app_one}.}
\label{fig:app_two}
\end{figure}

In the above discussion, the lab-frame coordinates are all defined with the positive z axis anti-parallel to the electron beam's direction. Using Eq. \eqref{frame_relation} then, allowed us to calculate a high-energy particle's lab kinematic quantities in terms of the kinematics in the co-linear frame $b$. This is fine in itself. However, an additional assumption was made that detector acceptance could be defined simply using this lab-frame coordinate system. While a realistic EIC detector will certainly have a symmetric acceptance around the electron beam direction for the electron end-cap, it is overwhelmingly likely that the detector acceptance in the hadron end-cap will be symmetric around the incoming proton beam direction. The positive acceptance limit will then be at an approximately fixed value of $\eta_{lab,prot}$, which is the pseudo-rapidity in the lab frame calculated relative to the proton beam direction, rather than at a fixed value of $\eta_{lab}$. The dependence of $\eta_{lab,prot}$ on $\eta_{gen}$ for fixed values of $\phi_{gen}$ is shown in figure~\ref{fig:app_two}. It is the mirror image of figure~\ref{fig:app_one}. So we can conclude that a detector that extends from $\eta_{lab} = -3.5$ to $\eta_{lab,prot} = +3.5$ is equivalent to an acceptance in the co-linear frame $b$ of $-3.5 < \eta_{gen} < +3.5$.

\clearpage

\section*{Acknowledgements}
We thank the members of the EICUG collaboration who initiated the process of the EIC Detector Design and the EIC Detector Yellow Report, through which this work began. We acknowledge the support of the Center for Frontiers in Nuclear Science (CFNS) (B.S \& A.P-L) and the DOE Award \# DE-FG02-05ER41372 (A.D). One of us (A.P-L) was partially supported by the GEM Fellowship at Brookhaven National Laboratory.

\bibliography{references}
\bibliographystyle{unsrt}

\end{document}